\documentclass[fleqn,usenatbib]{mnras}

\usepackage{newtxtext,newtxmath}

\usepackage[T1]{fontenc}

\DeclareRobustCommand{\VAN}[3]{#2}
\let\VANthebibliography\thebibliography
\def\thebibliography{\DeclareRobustCommand{\VAN}[3]{##3}\VANthebibliography}

\usepackage{graphicx}	
\usepackage{amsmath}	
\usepackage{longtable} 
\usepackage{booktabs} 
\usepackage{caption}
\usepackage{subcaption}

\title[The blazars TXS\,0506+056 and PKS\,0502+049]{The relativistic parsec-scale jets of the blazars TXS\,0506+056 and PKS\,0502+049 and their possible association with gamma-ray flares and neutrino production}

\author[V. Y. D. Sumida et al.]{
Viktor Y. D. Sumida,$^{1,2}$\thanks{E-mail: viktor.sumida@alumni.usp.br (VYDS)}
A. de A. Schutzer,$^{3}$
A. Caproni$^{1}$
and Z. Abraham$^{4}$
\\
$^{1}$N\'ucleo de Astrof\'\i sica, Universidade Cidade de S\~ao Paulo R. Galv\~ao Bueno 868, Liberdade, S\~ao Paulo, SP, 01506-000, Brazil\\
$^{2}$Universidade Cruzeiro do Sul, R. Galv\~ao Bueno 868, Liberdade, 01506-000, S\~ao Paulo, SP, Brazil\\
$^{3}$Univ. Grenoble Alpes, CNRS, IPAG, 38000 Grenoble, France\\
$^{4}$Instituto de Astronomia, Geof\'\i sica e Ci\^encias Atmosf\'ericas, Universidade de S\~ao Paulo, R. do Mat\~ao 1226, Cidade Universit\'aria,\\
	05508-900, S\~ao Paulo, SP, Brazil
}

\date{Accepted XXX. Received YYY; in original form ZZZ}

\pubyear{2021}

\begin{document}
\label{firstpage}
\pagerange{\pageref{firstpage}--\pageref{lastpage}}
\maketitle

\begin{abstract}
The physical nature of the mechanism responsible for the emission of neutrinos in active galactic nuclei (AGN) has been matter of debate in the literature, with relativistic jets of radio-loud AGNs as possible candidates to be the sources of high energy neutrinos. The most prominent candidate so far is the blazar TXS\,0506+056, which is found to be associated with the neutrino event IceCube-170922A. Furthermore, the IceCube reported an excess of neutrinos towards TXS\,0506+056 between September 2014 and March 2015, even though this association needs additional investigation, considering the presence of a nearby $\gamma$-ray source, the quasar PKS\,0502+049. Motivated by this, we studied the parsec-scale structures of TXS\,0506+056 and PKS\,0502+049 through radio interferometry at 8 and 15\,GHz. We identified twelve jet components in TXS\,0506+056 and seven components in PKS\,0502+049. The most reliable jet components show superluminal speeds ranging from $9.5c$ to $66c$ in the case of TXS\,0506+056, and from $14.3c$ to $59c$ for PKS\,0502+049, which were used to estimate a lower (upper) limit for the Lorentz factor (jet viewing angle) for both sources. A novel approach using simultaneously the brightness temperature of the core region and the apparent speeds of the jet components allowed us to infer basic jet parameters for TXS\,0506+056 at distinct epochs. We also found that the emergence of new jet components coincides with the occurrence of gamma-ray flares. Interestingly, two of these coincidences in the case of PKS\,0502+049 and one for TXS\,0506+056 seems to be correlated with neutrino events detected by the IceCube Observatory.
\end{abstract}

\begin{keywords}
BL Lacertae objects: individual (TXS\,0506+056) --- galaxies: active --- galaxies: jets --- neutrinos --- Quasars: individual (PKS\,0502+049) --- techniques: interferometric
\end{keywords}



\section{Introduction}\label{sec:intro}

Blazars are a subclass of active galaxies whose jets  form a small angle with respect to our line of sight and usually exhibit strong variability from radio up to TeV gamma-rays, presenting a non-thermal continuum and a relativistic jet \citep[e.g.,][]{Urry_Padoavani_1995}. They have also been considered  powerful cosmic particle accelerators and thus  prominent candidates of high-energy (HE) astrophysical neutrinos generated in hadronic interactions in their jets \citep[e.g.,][]{Stecker_et_al_1991, Mannheim_1995, Gao_et_al_2017, Murase_2017, Rodrigues_et_al_2018, Murase_et_al_2018}.

Located at a redshift of $0.3356$ \citep{Paiano_et_al_2018}, the BL Lac object\footnote{\citet{2019MNRAS.484L.104P} reclassified this source as a flat-spectrum radio quasar with hidden broad lines and a standard accretion disc or simply a ``masquerading BL Lac object".} TXS\,0506+056 was recently proposed as the object that produced the 290 TeV muon neutrino related to the IceCube-170922A detection \citep{IceCube_Collaboration_et_al_2018}. This event has opened new perspectives for investigating the multi-messenger physics of blazars jets. Moreover, motivated by this HE neutrino detection and after an archival search, the IceCube collaboration reported a 3.5\,$\sigma$ excess of neutrinos from the direction of TXS\,0506+056, with energy above 30 TeV, during a time window from September 2014 to March 2015 \citep{IceCube_Collaboration_2018}.
The interpretation of such events turned out to be a difficult task, since the source TXS\,0506+056 was in a quiescent state of both the radio and gamma-ray emission during this period \citep{Liang_et_al_2018, Padovani_et_al_2018}. 

The flat spectrum radio quasar (FSRQ) PKS\,0502+049 is separated by an angular distance of $1\fdg2$ from TXS\,0506+056 \citep{Johnston_et_al_1995J}.
At a redshift of 0.954 \citep{Drinkwater_et_al_1997}, PKS\,0502+049 was identified as a GeV gamma-ray emitter just before and immediately after the period of the neutrino excess in 2014-2015 \citep{He_et_al_2018}, so its contribution may not be disregarded. 

Different mechanisms responsible for the production of these detected neutrinos and gamma-rays from both blazars have been proposed in the literature \citep[e.g.,][]{Ansoldi_et_al_2018, He_et_al_2018, Sahakyan_2018, Sahakyan_2019, Rodrigues_et_al_2019, Liu_et_al_2019, 2019PhRvD..99j3006B, Banik_et_al_2020}. However, the results remain controversial.

Some groups have studied the kinematics of the parsec-scale jet of TXS\,0506+056 using the standard Difmap tasks \citep{Shepherd_1997} to model the sky brightness distribution for VLBA observations in the $(u,v)$ visibility plane.
\cite{Kun_et_al_2019}, \cite{Lister_et_al_2019}, and \cite{Li_et_al_2020} identified four peculiar features which maintain quasi-stationary separations from the core, in a region extending up to 4\,mas. On the other hand, \cite{Britzen_et_al_2019} suggested two scenarios in order to explain the evidence for a strong curvature in the jet of TXS\,0506+056. The first scenario is characterised by one strongly curved jet while the second  is defined by a structure made up of two jets. None of these studies could provide any temporal correlation between the ejection of the jet components and the HE neutrinos events. In the case of PKS\,0502+049, there is no published kinematic study from the best of our knowledge.

The correlated observation of HE neutrinos and enhanced gamma-ray activity with ejection of jet components may shed some light to the production scenario of these detected neutrinos and gamma-rays from the blazars. Thus, the primary objective of this work consists of a kinematic study of the parsec-scale jets of the blazars TXS\,0506+056 and PKS\,0502+049. We have gathered existing VLBI data of TXS\,0506+056 from the MOJAVE/VLBA Survey archive \citep{Lister_et_al_2009} to apply the global optimisation statistical technique Cross-Entropy \citep[hereafter CE;][]{Rubinstein_1997}. Adapted by \cite{Caproni_et_al_2011}, this method allows to model interferometric radio images of astrophysical jets and estimate a minimum number of discrete two-dimensional elliptical Gaussian components at each epoch \citep[e.g.,][]{Caproni_et_al_2014, Caproni_et_al_2017}. We also investigated kinematically whether the estimated ejection epochs of PKS\,0502+049 jet components, which accompany the observed gamma-ray flux, may or may not contribute to the events detected by the IceCube Observatory.

This work is structured as follows: the observational data set used in this work is presented in \autoref{sec:ObsData}. The general results from our kinematic studies of the parsec-scale jets of TXS\,0506+056 and PKS\,0502+049, including a novel approach to estimate the values of some jet parameters are shown in \autoref{sec:Results}. A possible connection between jet components and neutrino emission are explored in \autoref{sec:Disc}. Final remarks are presented in \autoref{sec:concl}. We assume throughout this work $H_0$ = 71 km s$^{-1}$ Mpc$^{-1}$, $\Omega_\mathrm{M}$ = 0.27 and $\Omega_\Lambda$ = 0.73, implying that 1.0 mas = 4.78 pc and 1.0 mas yr$^{-1}$ = 20.84$c$ for TXS\,0506+056, where $c$ is the speed of light. In the case of PKS\,0502+049, 1.0 mas = 7.94 pc and 1.0 mas yr$^{-1}$ = 50.63$c$.

\section{Data analysis}\label{sec:ObsData}

\subsection{Radio interferometric data}\label{sec:ObsDataRadio}

We investigate the structure of the parsec-scale jet of TXS\,0506+056 and PKS\,0502+049 by employing available VLBI data from archival databases. These data consist of naturally weighted total intensity maps obtained at 15\,GHz, publicly available  at MOJAVE/2-cm Survey Data Archive\footnote{\url{http://www.physics.purdue.edu/astro/MOJAVE/index.html}} \citep{Lister_et_al_2009}, spanning   from 2009 to 2020 and 2016 to 2020 for TXS\,0506+056 and PKS\,0502+049, respectively. One additional map for PKS\,0502+049 was acquired from the Astrogeo Center\footnote{\url{http://astrogeo.org/vlbi_images/}} at 8\,GHz in November 2018. In \autoref{table:Characteristics_TXS_table} and \autoref{table:Characteristics_PKS_table}, we list the main characteristics of these images for TXS\,0506+056 and PKS\,0502+049, respectively, represented by the peak intensity ($I_\mathrm{max}$), the root mean square (RMS) of the off-source surface brightness, and the parameters of the synthesised elliptical \textsc{clean} beam, which are the FWHM major axis ($\Theta^{\mathrm{FWHM}}_{\mathrm{beam}}$), eccentricity ($\epsilon_\mathrm{\,beam}$) and position angle ($\theta_\mathrm{beam}$) on the plane of the sky.

The FITS images are constituted by an array of $1024 \times 1024$ or $512 \times 512$ pixels in right ascension and declination directions. For the purpose of minimising the computational time required for our model-fitting algorithm to find the optimal solutions, we cropped the original data to maintain only the fraction with a useful signal (source) without compromising the obtained results \footnote{We show in \autoref{fig:comparison_CE} the results from the application of our CE model fitting to four cropped images (chosen randomly from the 37 maps analysed in this work) but doubling their sizes. Similar to \citet{Caproni_et_al_2011} and \citet{Caproni_et_al_2014}, no substantial differences (smaller than their associated error) were found among structural parameters of the jet components in relation to those reported in this section.}.

The radio maps of TXS\,0506+056 and PKS\,0502+049 reveal a relatively intense radio core and an inconspicuous jet that extends up to 4 mas on one side of the core. The surface brightness of these jets was decomposed in elliptical Gaussian components and their structural parameters were determined via the Cross-Entropy (CE) global optimisation technique \citep[e.g.,][]{Rubinstein_1997, Caproni_et_al_2014, Caproni_et_al_2017}. Each elliptical Gaussian component is characterised by six structural parameters: two-dimensional peak position ($x_0, y_0$), with coordinates $x$ and $y$ oriented in right ascension and declination directions, respectively; peak intensity, $I_0$, semi-major axis, $a$, eccentricity, $\epsilon=\sqrt{1-(b/a)^2}$, where $b$ is the semi-minor axis, and the position angle of the major axis, $\psi$, measured positively from west to north.

\begin{figure*}
	\centering
	\includegraphics[width=1\textwidth]{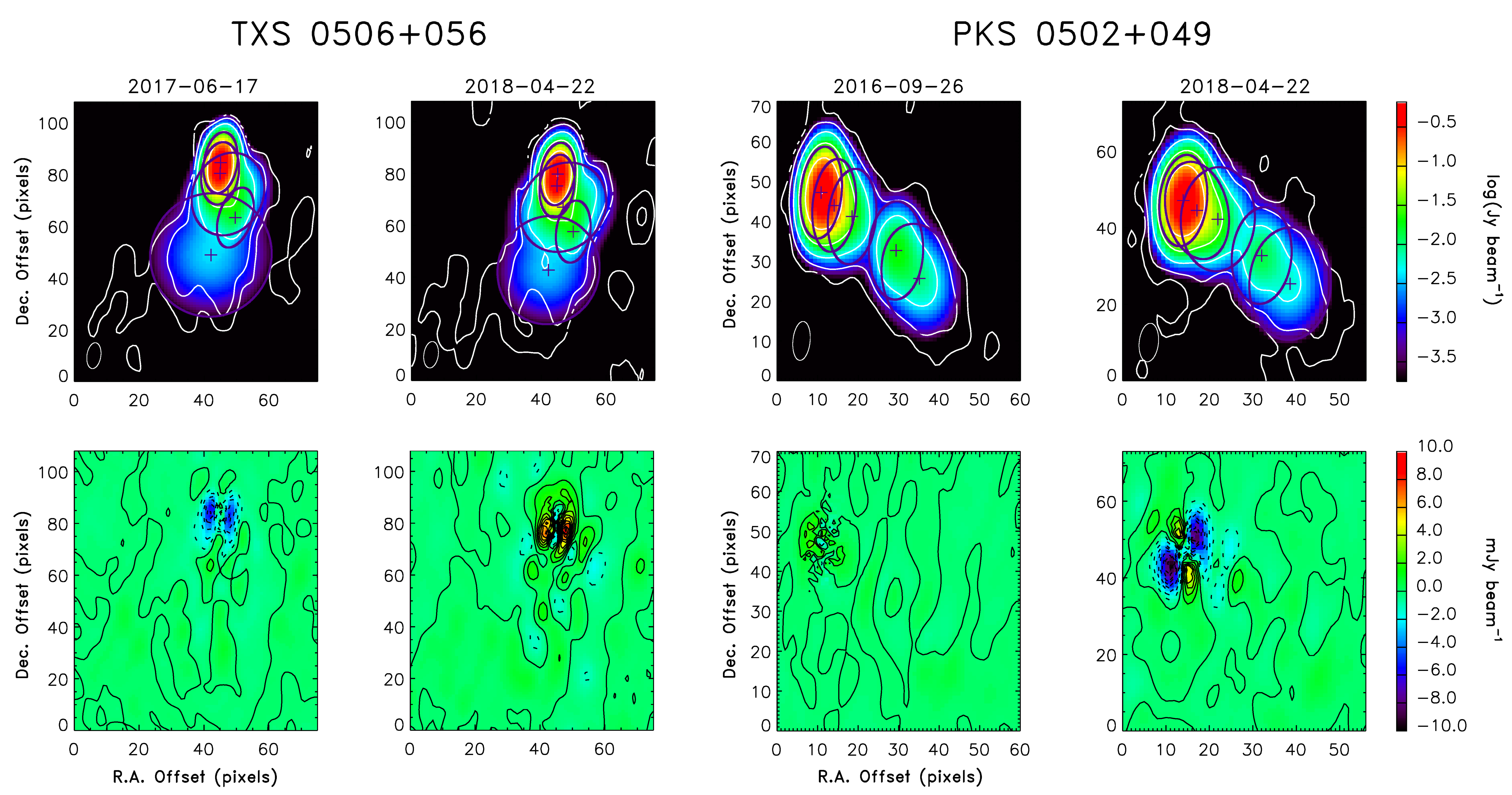}
    \caption{Parsec-scale radio structure of TXS\,0506+056 (left panels) and PKS\,0502+049 (right panels) at 15 GHz analysed in this work and their respective residual maps. 
    White contour lines depict the VLBA map, while the superimposed dark thick ellipses are the contours of the Gaussian CE model-fitting components at the FWHM (centre marked with crosses).
    The observation epochs are marked in the top of each panel and refer from left to right, respectively, to: 2017-06-17 - two months before the IceCube-170922A event; 2018-04-22 - seven months after the IceCube-170922A event; 2016-09-26 - corresponding to the first observation epoch of the available archival VLBA data of PKS\,0502+049, 1.5 yr after the 2014-2015 neutrino excess; 2018-04-22 - seven months after the IceCube-170922A event. The ellipse in the lower-left corner of the individual panels represents the FWHM of the elliptical synthesised \textsc{clean} beam. 
    }
    \label{fig:Gaussians_figure}
\end{figure*}

Following the criteria proposed in \cite{Caproni_et_al_2014}, we found the optimal number of Gaussian components in each image, leading to a minimum of two and a maximum of six components in the images analysed in this work. 
The CE model-fitting results of TXS\,0506+056 and PKS\,0502+049 on four representative epochs at 15\,GHz can be seen in \autoref{fig:Gaussians_figure}, in which the Gaussian components are shown superimposed on the observed image, as well as the respective residual maps.
These epochs were not chosen at random, we selected interferometric radio observations whose dates preceded or followed the IceCube-170922A event or the 2014-2015 neutrino excess.

The structural parameters of the Gaussian components derived by our CE optimisations are listed in Tables \ref{table:Parameters_TXS_table}, \ref{table:Parameters_PKS_15GHz_table} and \ref{table:Parameters_PKS_8GHz_table}. The flux density, $F$, (the entries in the third column of these tables) was estimated from
\begin{equation}
    F = 8 \ln{2} \left[ \frac{a^2 \sqrt{1 - \epsilon^2}}{\left(\Theta^\mathrm{FWHM}_{\mathrm{beam}}\right)^2 \sqrt{1 - \epsilon^2_{\mathrm{beam}}}} \right] I_0 \;.
\end{equation}

The formal uncertainties for the elliptical Gaussian parameters were obtained from weighted mean and standard deviation of the best tentative solution at each iteration during the CE optimisations (see Equations 10 and 11 in \citealt{Caproni_et_al_2011}). As pointed out in \citet{Caproni_et_al_2011}, these error estimates should be assumed as a lower limit for the true uncertainties of the derived Gaussian parameters. In the case of the uncertainties in the Gaussian peak positions, we also added a term corresponding to 1/10 of the FWHM restoring beam dimensions (e.g., \citealt{2009AJ....138.1874L}) in quadrature to the respective CE uncertainties in those quantities (e.g., \citealt{Caproni_et_al_2014}), providing more conservative estimates of the errors in the sky positions of the Gaussian components.

The last column of Tables \ref{table:Parameters_TXS_table}, \ref{table:Parameters_PKS_15GHz_table} and \ref{table:Parameters_PKS_8GHz_table} indicates the brightness temperature of the core in the observer’s reference frame, which will be discussed in detail in \autoref{sec:BrightTemp}.

The jet inlet region, or simply the core component, is the most intense component in terms of flux density found in our CE modellings for both sources, and it is always located at the northernmost part of the analysed interferometric images. The same components detected in the 8-GHz map of PKS 0502+049 are seen in the quasi-contemporaneous image at 15\,GHz, reinforcing the robustness of our CE modelling.

\subsection{Gamma-ray data}\label{sec:ObsDataGamma}

The $\gamma$-ray data analysed in this work was extracted from the public Fermi Large Area Telescope Pass 8 database \citep{2009ApJ...697.1071A, 2013arXiv1303.3514A}. A standard Fermi-LAT unbinned likelihood analysis\footnote{\url{ https://fermi.gsfc.nasa.gov/ssc/data/analysis/scitools/likelihoodtutorial.html}} was performed in order to obtain the light curves of PKS\,0502+049 and TXS\,0506+056 and unveil the association between the emergence of parsec-scale radio components and $\gamma$-ray flares, and also how these light curves are connected to the neutrino events reported by the IceCube Collaboration \citep{IceCube_Collaboration_2018}. 

The analysed data-set was extracted in the period of the 4th of August 2008 and the 05th of June 2019 and the respective photon energy range was between $300$ MeV$\leq E\leq 300$ GeV. The $\gamma$-ray data were selected in a circular Region Of Interest (ROI) with a radius of $6\degr$ centered on the TXS 0506+056. In order to associate the photons and fit the data to its respective objects, we have made use of the models for the source (iso\_P8R3\_SOURCE\_V2\_v1) and the diffuse background (gll\_iem\_v07) available in the Fermi-LAT FL8Y\footnote{\url{https://fermi.gsfc.nasa.gov/ssc/data/access/lat/fl8y/}}. After obtaining the light curve, the data were binned in intervals of 10 days for both sources in order to take into account weekly variations and not just monthly such as presented in other previous works \citep[e.g.,][]{IceCube_Collaboration_2018,Padovani_et_al_2018}. Moreover, in the bins that no photon was associated to the sources, the flux was therefore presented as zero. The resulting light curves are presented and discussed in the context of our kinematic studies of the parsec-scale jets of PKS\,0502+049 and TXS\,0506+056 in \autoref{sec:kin}.

\begin{figure*}
    \centering
     \begin{subfigure}{1.0\columnwidth}
         \includegraphics[width=\textwidth]{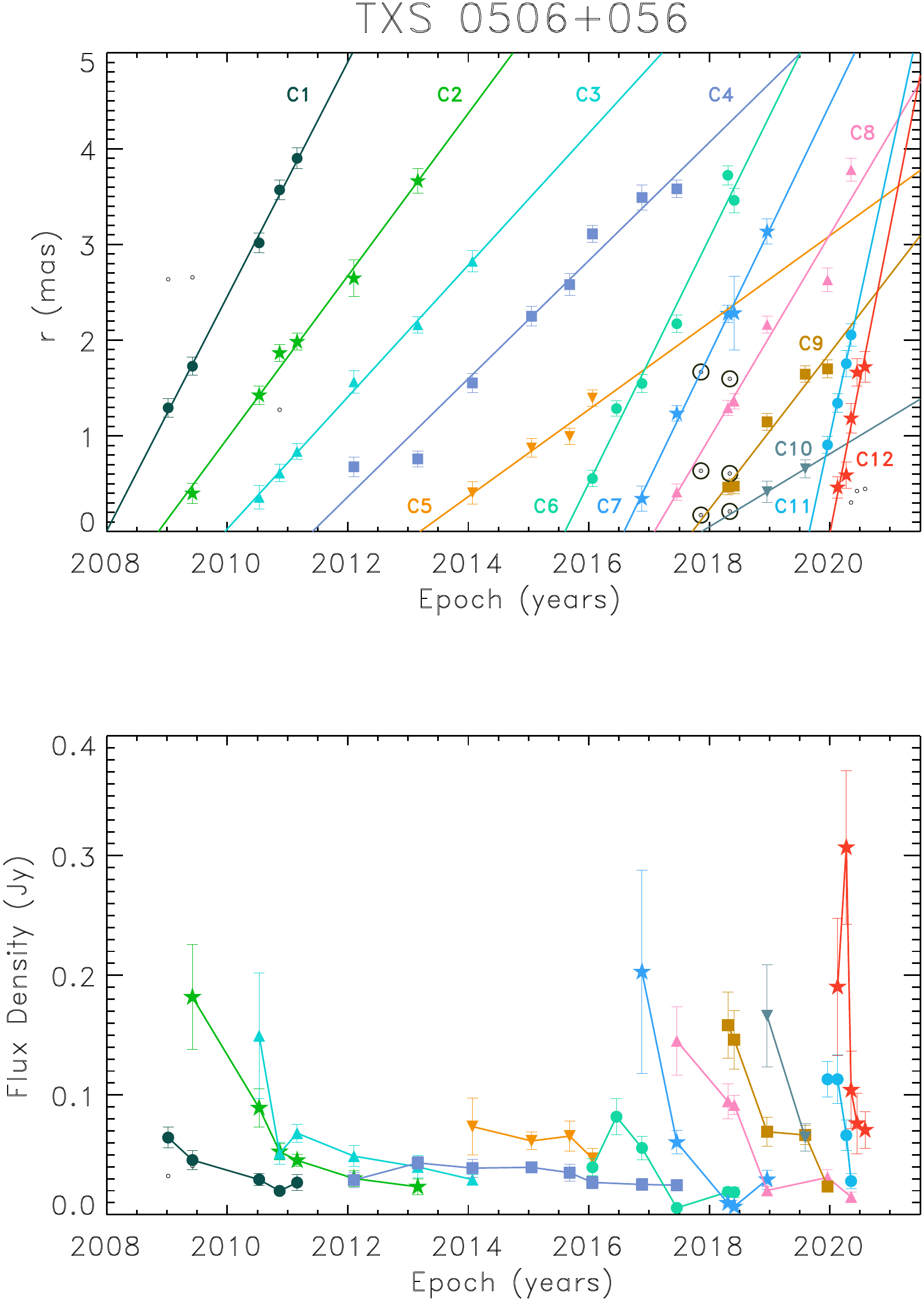}
         \label{fig:TXS0506_fitting_figure}
     \end{subfigure}
     \hfill
     \begin{subfigure}{1.0\columnwidth}
         \includegraphics[width=\textwidth]{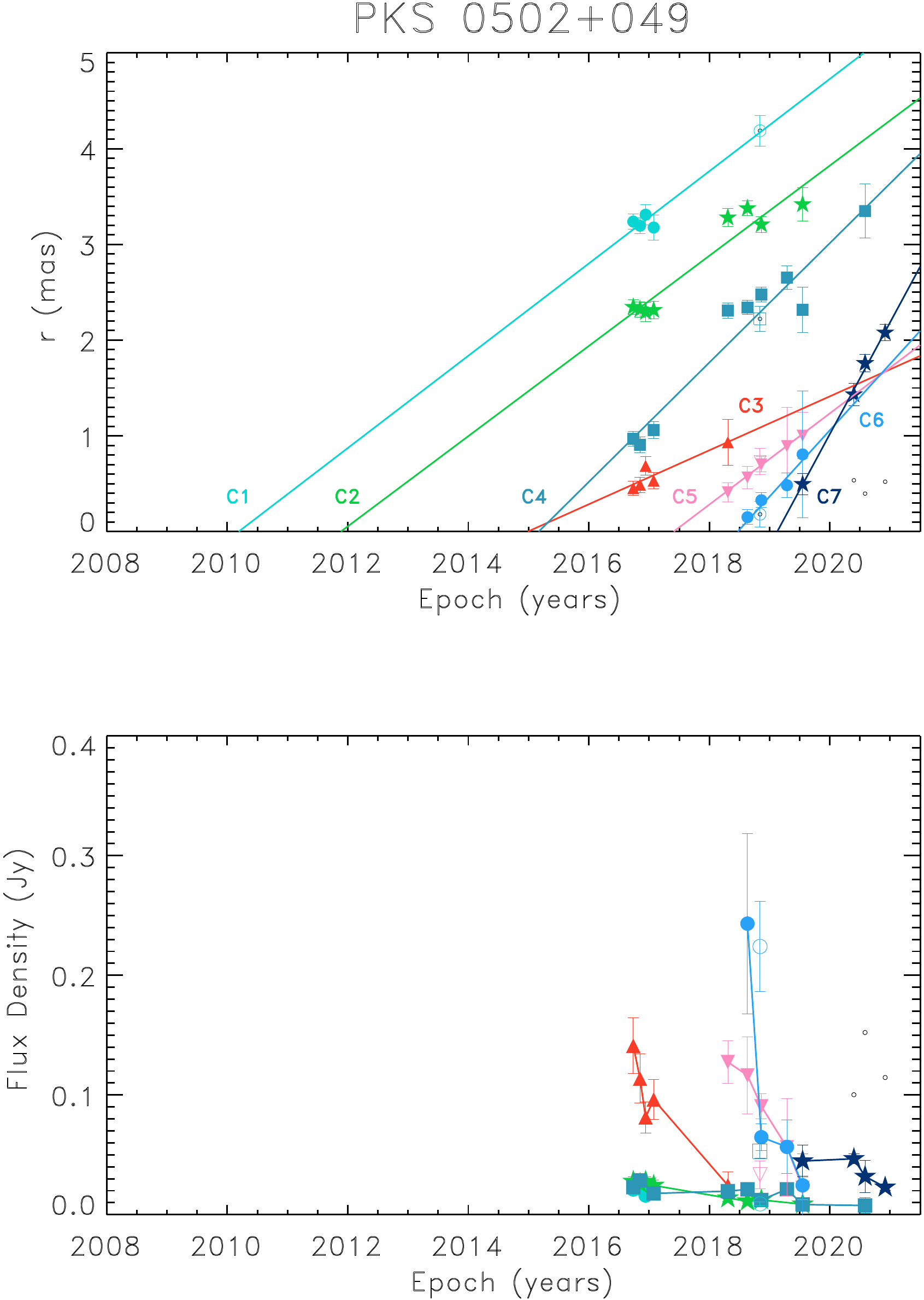}
         \label{fig:PKS0502_fitting_figure}
     \end{subfigure}
     \hfill
     \caption{Top panels: angular separation from the core of the jet components in TXS\,0506+056 (left panel) and PKS\,0502+049 (right panel) as a function of time. Solid symbols correspond to 15\,GHz observations data derived from multi-epoch observations. Coloured open symbols and black circles represent, respectively, the identified jet components in PKS\,0502+049 at 8\,GHz and those components identified in TXS\,0506+056 by \citet{Ros_et_al_2020} at 43\,GHz. Unidentified components are shown by black points The straight lines represent linear regressions fitting for the individual jet components. Bottom panels: flux density evolution of the jet components as a function of time for both sources.}
    \label{fig:TXS_PKS_fitting_figure}
\end{figure*}

\section{Results}\label{sec:Results}

\subsection{Kinematics of the jets components}\label{sec:kin}

The kinematic analysis at parsec scales requires the identification of jet components across consecutive epochs in our data set. Based on \cite{Caproni_et_al_2014}, we assumed a kinematic scenario for the parsec-scale jet whose individual components recede ballistically or quasi-ballistically from the stationary core (constant proper motions along straight trajectories on the plane of sky). Thus, we were able to trace their motions studying the temporal evolution of flux densities and position angles, as well as their distances relative to the (stationary) core. 

Thereby, the top panels in \autoref{fig:TXS_PKS_fitting_figure} show the time evolution of the core-component distance for each of the jet components identified in TXS\,0506+056 and PKS\,0502+049. The solid lines and the corresponding slopes represent their apparent proper motion ($\mu$) obtained by linear regression of the equation 

\begin{equation}
	r(t) = \mu\left(t-t_0\right) \; ,
	\label{eq:motionequation}
\end{equation}
\\where $r$ is the core-component distance at the instant $t$, and $t_0$ is the ejection epoch of the jet component ($r(t_0)=0$).

In the top left panel of \autoref{fig:TXS_PKS_fitting_figure}, the black circles represent the jet features identified in TXS\,0506+056 at 43 GHz by \cite{Ros_et_al_2020}. It is interesting to note that all of these 43-GHz features can be associated with the jet components C7, C8, C9 and C10 identified in the present work.

Based on the broadly accepted shock-in-jet model \citep[e.g.,][]{Konigl_1981, Marscher_Gear_1985, Hughes_et_al_1985, Valtaoja_et_al_1992, Turler_et_al_2000},
jet components represent shock waves propagating downward in a relativistic jet and go through three distinct stages: an initial stage dominated by inverse Compton cooling, 
a synchrotron-dominated loss stage, and a final phase where the losses are adiabatic.
We can see from bottom panels in \autoref{fig:TXS_PKS_fitting_figure} the time evolution of flux density of individual components identified in both blazars. 

We list the kinematic parameters of each of these jet components for TXS\,0506+056 and PKS\,0502+049 in \autoref{table:Kin_Param_TXS_PKS_table}. The parameter 
$\beta_{\mathrm{app}}$ correspond
to 
the apparent speed in units of $c$, while $\bar{\eta}$ represents the mean position angle of a jet component among $N_{\mathrm{epoch}}$ epochs used in the fit. Uncertainties for $t_{0}$ and $\mu$ were calculated from the linear fit, taking into account the errors on the individual positions of the jet components, while standard deviation of $\eta$ was used as an error estimator for $\bar{\eta}$. 
The numbers displayed in the last column give the probability $p$ of the chi-squared value to be less than or equal to the value obtained from the linear regressions presented in \autoref{fig:TXS_PKS_fitting_figure}.

\begin{table*}
	\centering
	\parbox{12.5cm}{\caption{Kinematic parameters of the CE model-fitting jet components of TXS\,0506+056 and PKS\,0502+049 identified in this work.} 	\label{table:Kin_Param_TXS_PKS_table}} 
	\begin{tabular}{ccccccc}
		\toprule
		Jet component &  $t_{0}$ & $\mu$ & $\beta_{\mathrm{app}}$ & $\bar{\eta}$ & $N_{\mathrm{epoch}}$ $^\mathrm{a}$ & $p^{\mathrm{b}}$\\ & (yr) & (mas\,yr$^{-1}$) & & (deg) & & \\
		\hline 
		\multicolumn{7}{c}{TXS\,0506+056}\\
		\hline
  C1 & 2007.99  $\pm$  0.08 &  1.22  $\pm$   0.12 &   25.5  $\pm$    2.6 &  -182.5  $\pm$     0.1      &  5      &  0.028 \\
  C2 & 2008.86  $\pm$  0.11 &  0.85  $\pm$   0.08 &   17.7  $\pm$    1.7 &  -177.3  $\pm$     0.5      &  6      &  0.066 \\
  C3 & 2009.96  $\pm$  0.14 &  0.69  $\pm$   0.07 &   14.4  $\pm$    1.5 &  -173.3  $\pm$     0.4      &  6      &  0.006 \\
  C4 & 2011.41  $\pm$  0.13 &  0.62  $\pm$   0.05 &   12.9  $\pm$    1.0 &  -179.0  $\pm$     0.1      &  8      &  0.472 \\
  C5 & 2013.20  $\pm$  0.3 &  0.45  $\pm$   0.15 &    9.5  $\pm$    3.2 &  -168.1  $\pm$     0.3      &  4      &  0.235 \\
  C6 & 2015.60  $\pm$  0.07 &  1.28  $\pm$   0.11 &   26.7  $\pm$    2.2 &  -177.8  $\pm$     0.2      &  6      &  0.492 \\
  C7 & 2016.58  $\pm$  0.08 &  1.31  $\pm$   0.14 &   27.2  $\pm$    2.9 &  -175.0  $\pm$     0.9      &  5      &  0.069 \\
  C8 & 2017.09  $\pm$  0.09 &  1.06  $\pm$   0.09 &   22.2  $\pm$    2.0 &  -180.7  $\pm$     0.3      &  6      &  0.789 \\
  C9 & 2017.72  $\pm$  0.13 &  0.82  $\pm$   0.16 &   17.0  $\pm$    3.3 &  -174.1  $\pm$     0.2      &  5      &  0.220 \\
 C10 & 2017.9  $\pm$  0.4 &  0.4  $\pm$   0.5 &    8  $\pm$   11 &  -179.9  $\pm$     0.2      &  2      &  1.000 \\
 C11 & 2019.66  $\pm$  0.04 &  2.9   $\pm$    0.8 &   60  $\pm$   16 &  -176.4  $\pm$     0.5      &  4      &  0.034 \\
 C12 & 2020.01  $\pm$  0.03 &  3.2   $\pm$    0.7 &   66  $\pm$   14 &  -181.2  $\pm$     1.3      &  5      &  0.590 \\
        \hline 
		\multicolumn{7}{c}{PKS\,0502+049}\\
		\hline
C1 & 2010.2  $\pm$  0.5 &  0.48   $\pm$   0.13 &   24.4  $\pm$    6.6 &  -132.7  $\pm$     0.6      &  5      &  0.100 \\
C2 & 2011.9  $\pm$  0.3 &  0.47   $\pm$   0.08 &   23.9  $\pm$    4.0 &  -130.0  $\pm$     0.7      &  8      &  0.235 \\
C3 & 2015.0  $\pm$  0.4 &  0.28   $\pm$   0.18 &   14.3  $\pm$    9.1 &  -130.8  $\pm$     5.9      &  5      &  0.072 \\
C4 & 2015.17  $\pm$  0.12 &  0.62   $\pm$   0.06 &   31.6  $\pm$    3.1 &  -129.0  $\pm$     2.1      & 10      &  0.536 \\
C5 & 2017.4  $\pm$  0.2 &  0.5   $\pm$   0.2 &   24  $\pm$   12 &  -130  $\pm$     9      &  6      &  <0.001 \\
C6 & 2018.47  $\pm$  0.17 &  0.7   $\pm$   0.3 &   35  $\pm$   16 &  -120  $\pm$    24      &  4      &  0.058 \\
C7 & 2019.13  $\pm$  0.19 &  1.2    $\pm$   0.2  &   59  $\pm$   12 &  -127.3  $\pm$     0.6      &  4      &  0.057 \\
		\bottomrule
	\end{tabular}
	\parbox{12.5cm}{\flushleft $^\mathrm{a}$ Number of epochs for which a given jet component was detected by our CE model fitting.\\
	$^\mathrm{b}$ Probability that a chi-squared value is less than or equal to the value obtained in our linear regressions.}
\end{table*}


In the case of TXS\,0506+056, the linear fit for the jet components C1, C2, C3, C7 and C11 exhibited the highest confidence levels among all the identified jet components ($p \leq 0.1$), while C10 has a questionable fitting if we adopt the usual criterion $p\le 0.9$ (or similarly $q=1-p\ge 0.1$ for normal-distributed and well-estimated uncertainties; e.g., \citealt{1992nrca.book.....P, 10.5555/1965575}) as a discriminator of the reliability of a fit. It was already expected since C10 was only detected in two epochs. In addition, correlation coefficient $r_\mathrm{corr}$ (e.g., \citealt{cohen1988spa, 1992nrca.book.....P, heumann2017introduction, 2021NewA...8801602G}) of these linear regressions is always larger than about 0.94, which can be considered as an indicative of reasonable fits. Even though these statistical tests argue in favour of the significance of the majority of those linear fits, the ad-hoc ballistic-motion assumption might have introduced some bias in such analyses in the sense that it was incorporated in the process of the kinematic identification of the individual jet components. Thus, both statistical tests used to quantify the goodness of the fits do not exclude permanently more complex kinematic scenarios for the parsec-scale jet of TXS\,0506+056.

The core-component distance plot in  \autoref{fig:TXS_PKS_fitting_figure} shows a clear superposition between proper motion extrapolation of C5 (orange line) and the distance of C6 from the core between 2016.5 and 2017.5. The detection of C6 in 2016.1 independent of the number of Gaussian components assumed in our CE optimisations, as well as the difference of about 10 degrees between the mean position angles of C5 and C6 suggests the latter is real. In addition, the peak-like behaviour seen in the light curve of C6  (\autoref{fig:TXS_PKS_fitting_figure}) could be indicating a possible superposition between jet components C5 and C6 between 2016.5 and 2017.5, even though our CE optimisations were not able to detected (no splitting of C6 is found after increasing the number of components in our CE model fittings in this interval).

\begin{figure*}
     \begin{subfigure}{\textwidth}
       \centering         \includegraphics[width=0.75\textwidth]{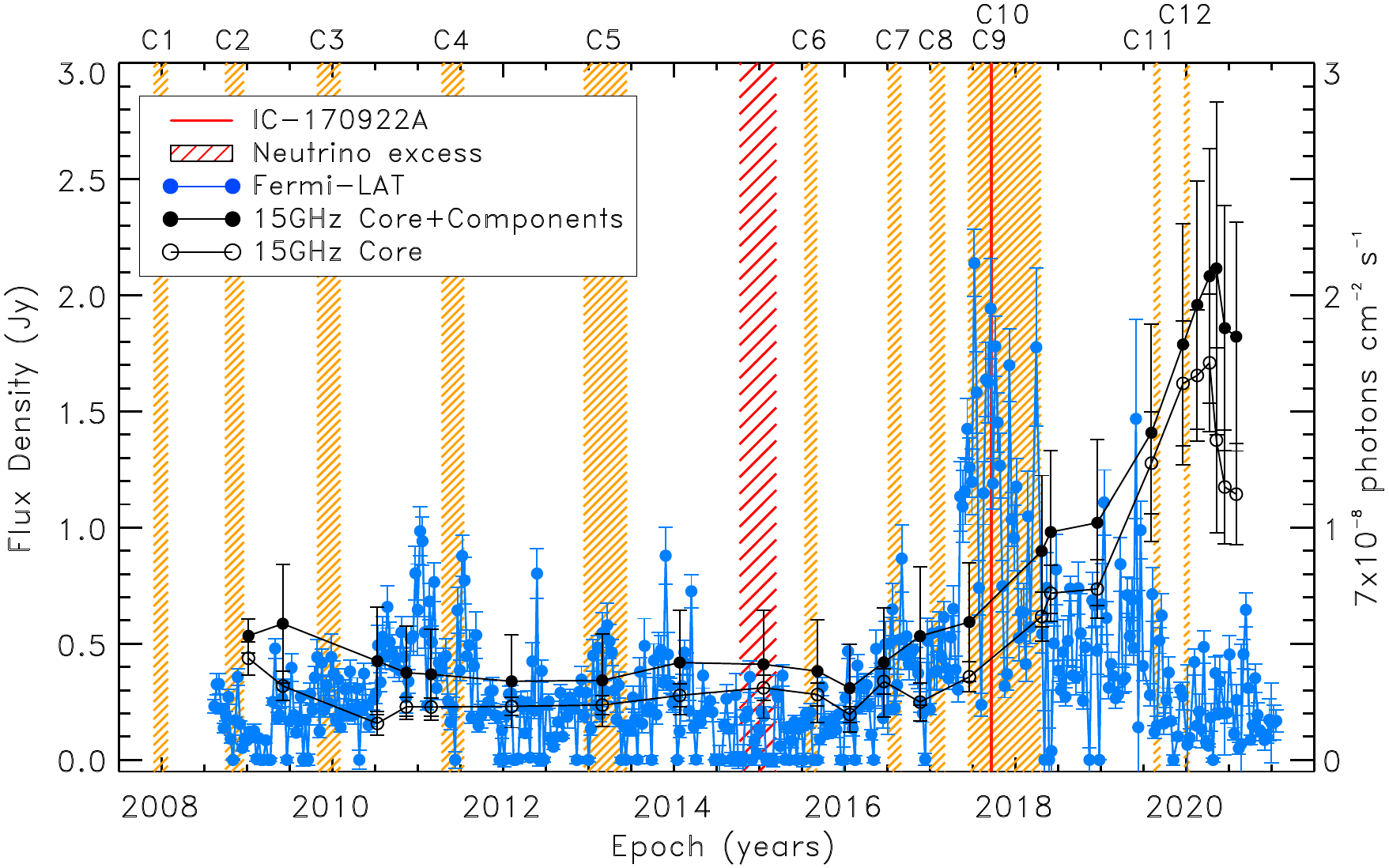}
     \end{subfigure}
     \hfill
     \begin{subfigure}{\textwidth}
       \centering         \includegraphics[width=0.75\textwidth]{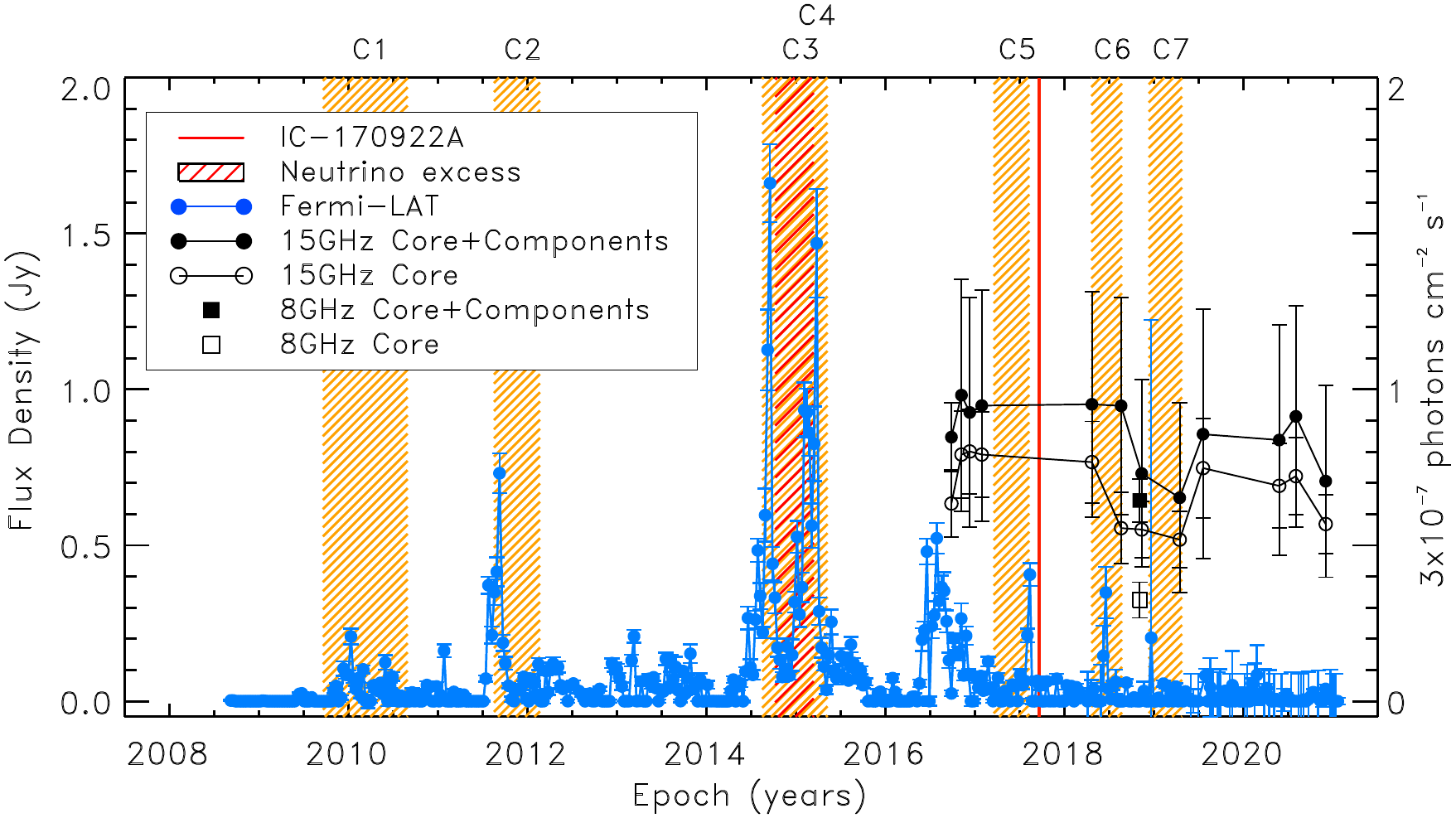}
     \end{subfigure}
     \hfill
     \caption{Flux density behaviour of TXS\,0506+056 (top panel) and PKS\,0502+049 (bottom panel). Solid and open black circles display the total (core plus jet components) and core flux densities at 15\,GHz, respectively. Solid and open black square refer, respectively, to the total and core flux densities for the single 8-GHz observational epoch of PKS\,0502+049. Fermi gamma-ray light curve is shown by blue circles. Vertical orange bars highlight the 1$\sigma$ uncertainty range for the ejection epochs of the jet components.}
    \label{fig:TXS_PKS_flux_figure}
\end{figure*}

Good statistical reliability was found for five out of seven jet components identified in PKS\,0502+049 ($p \la 0.1$ and $r_\mathrm{corr}\ga 0.92$). It is important to note that the lack of VLBI monitoring of PKS\,0502+049 between 2017 and 2018 casts some doubts on the kinematic identifications of the jet components C1 and C2 shown in \autoref{fig:TXS_PKS_fitting_figure}. Even though the present data cannot rule out a scenario where C1 and C2 are standing or slowly moving jet features (as observed in many other blazars), their inferred emergence epochs coincide with a flaring state of the source at gamma-ray energies as seen in \autoref{fig:TXS_PKS_flux_figure}, providing extra support for our C1 and C2 identifications. Whether or not C1 and C2 are quasi-stationary components does not exert any influence on the results presented in the next sections of this work. The identification of the jet component C3 might also be impacted by the same one-year data gap mentioned previously. However, the occurrence of strong gamma-ray flares during the emergence of C3 argues in favour of its possible existence.

Since time-correlation between the radio and gamma-ray activity has been found in other AGNs \citep[e.g.,][]{Max-Moerbeck_et_al_2014, Richards_et_al_2014}, as well as the relationship between ejection of new jet components and occurrence of flares in gamma-rays \citep[e.g.,][]{Otterbein_et_al_1998, Jorstad_et_al_2001, Agudo_et_al_2011, Cutini_et_al_2014, Lisakov_et_al_2017}, we looked for similar correlations in the case of TXS\,0506+056 and PKS\,0502+049. We plot 
in \autoref{fig:TXS_PKS_flux_figure} 
the public data from the \textit{Fermi}-LAT (see details in \autoref{sec:ObsDataGamma}), as well as the time behaviour of the parsec-scale core and total flux densities (the sum of the contributions from core and individual jet components). Vertical orange bars in this figure represent the 1$\sigma$-uncertainty range for the ejection epochs of the jet components. Except for component C1 in TXS\,0506+056, for which no gamma-ray data is available at its ejection epoch, the remaining jet components have emergence epochs that coincide with enhanced activity seen in the $\textit{Fermi}$-LAT gamma-ray light curve. 

The radio flux density variability of the core plus jet components seen in \autoref{fig:TXS_PKS_flux_figure} is very similar to that obtained by \cite{Kun_et_al_2019} for TXS\,0506+056; these authors reported an increase of radio emission after the neutrino excess. Interestingly, in the period following  the HE neutrino event IC-170922A, the observed radio flux roughly increased a factor of 4, being accompanied by the appearance of 
two new superluminal components (C11, C12).
Our kinematic-based identification reveals that no component was ejected in TXS\,0506+056 during the six-month period in 2014-2015 (see \autoref{fig:TXS_PKS_flux_figure}), corroborating the temporarily quiescent state of this source in GeV emission.

Perhaps, one of the most important results in this work is  the identification of the jet component C9, in TXS\,0506+056, and C3 and C4, in PKS\,0502+049. 
As already mentioned, PKS\,0502+049 was in a state of enhanced gamma-ray emission just before and after the period of the neutrino excess. The jet component C3, whose apparent speed is $\beta_{\mathrm{app}} = 14.3c$, has an estimated ejection time of $2014.99 \pm  0.36$, coinciding with 
the 2014 gamma-ray flare and the neutrino excess. PKS\,0502+049 also ejected a very bright moving feature, C4, in $2015.17 \pm 0.12$, with an apparent speed of $31.6c$ in temporal coincidence with 2015 gamma-ray flare at $1\sigma$-level. The aforementioned events and their correlations can be seen in the bottom panel of \autoref{fig:TXS_PKS_flux_figure}.

The direction of the reported event IceCube-170922A was consistent with the location of TXS\,0506+056, which was observed in enhanced gamma-ray activity by \textit{Fermi}-LAT. Corroborating the association of HE neutrino with TXS\,0506+056, the jet component C9 identified in this work has an ejection epoch that coincides, within the uncertainties, with the event IC-170922A and the $\gamma$-ray flare that occurred on 22 September, 2017.

\begin{figure*}
     \begin{subfigure}{1.0\columnwidth}
       \centering          \includegraphics[width=0.70\textwidth]{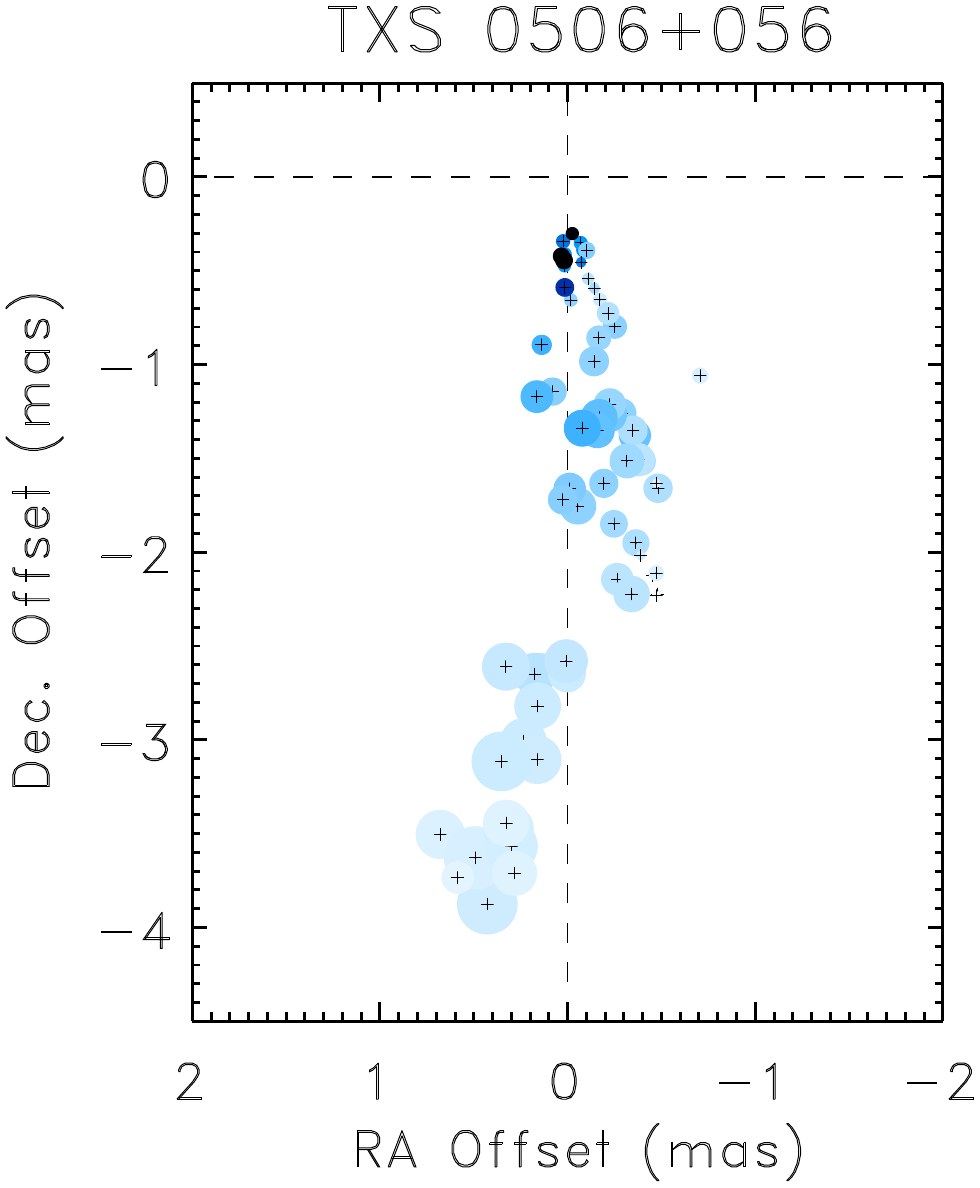}
     \end{subfigure}
     \hfill
     \begin{subfigure}{1.0\columnwidth}
       \centering         \includegraphics[width=0.70\textwidth]{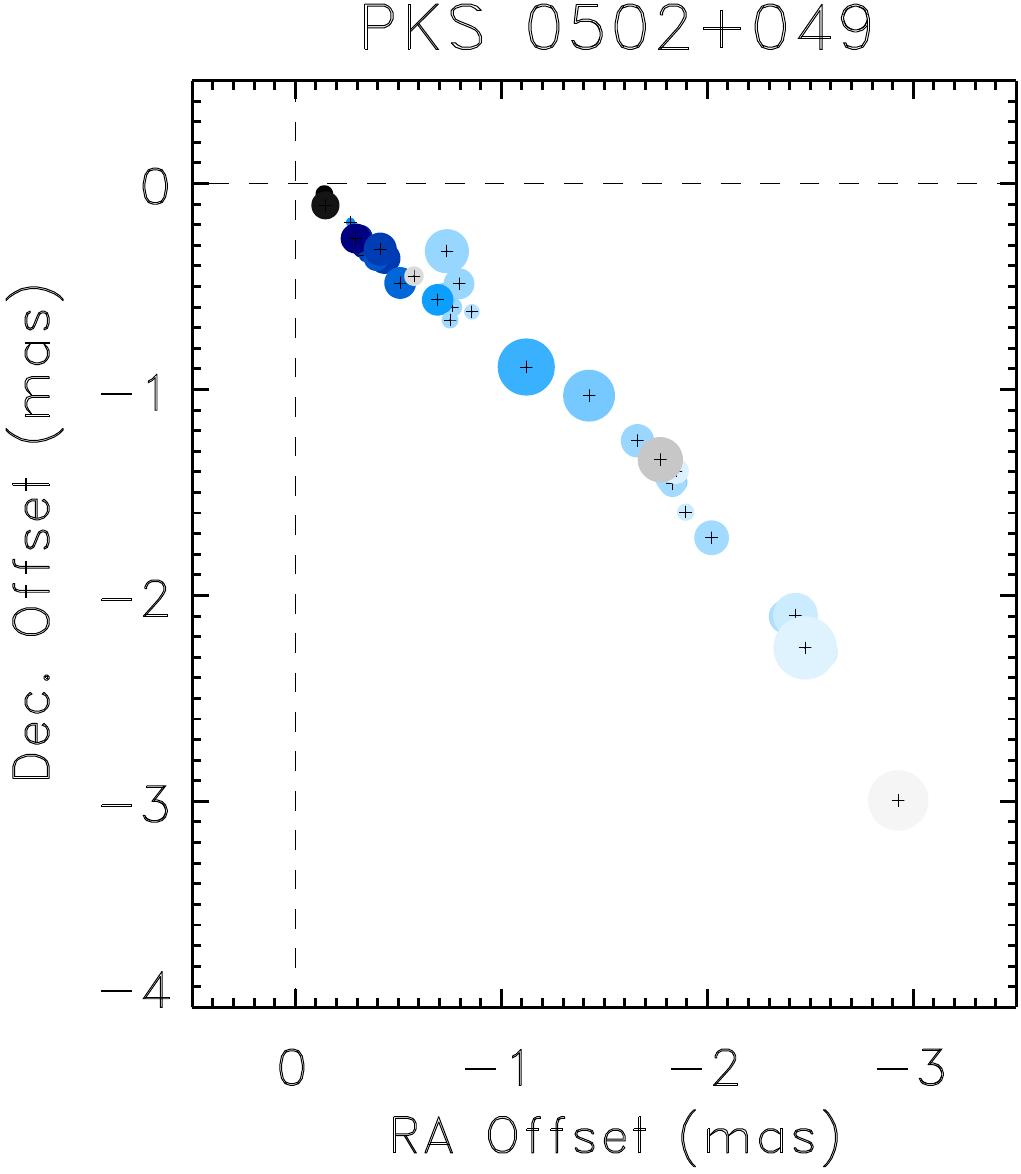}
     \end{subfigure}
     \hfill
     \caption{The distribution of right ascension and declination offsets of the jet components in TXS\,0506+056 (left panel) and PKS\,0502+049 (right panel) at 8\,GHz (grey circle) and 15\,GHz (blue circle). The size of the symbols is proportional to the semi-major axis of the jet components, while the shades of the colors are proportional to the jet component’s brightness, namely darker colors represent brighter components.}
    \label{fig:TXS_PKS_RA_Dec_Distribution_figure}
\end{figure*}

The spatial distribution of right ascension and declination offsets of the jet components relative to the position of the VLBI core is shown in \autoref{fig:TXS_PKS_RA_Dec_Distribution_figure}. TXS\,0506+056 has a jet-component distribution that extends approximately in a direction close to North-South. A more pronounced and systematic jet bending occurs at a core distance of $\sim2.5$ mas, which was interpreted in the framework of two distinct scenarios: a helical jet structure \citep{Kun_et_al_2019} and two relativistic jets in collision \citep{Britzen_et_al_2019}. \citet{Ros_et_al_2020} analysed two 43-GHz images of the radio jet of TXS\,0506+056 and concluded that they do not show any clear isolated knots of locally enhanced brightness temperature that could be associated with a secondary jet inlet region located $\sim1.2$ mas apart from the (primary) core, as suggested by \cite{Britzen_et_al_2019}.
Similarly to \cite{Ros_et_al_2020}, we have not identified any bright and persistent feature at the same region that could be attributed to the origin of a secondary jet in TXS\,0506+056.

The wider jet-component distribution along the right ascension direction for a given declination offset in TXS\,0506+056 suggests that jet orientation could have changed during the VLBI monitoring period analysed in this work. Moreover, the variations in the individual apparent speeds and position angles of the jet components shown, e.g., in \autoref{table:Kin_Param_TXS_PKS_table} seems to corroborate such possibility. Jet precession is one of the physical mechanisms that could drive those temporal changes. Indeed, \cite{Britzen_et_al_2019} constructed a precession model for TXS\,0506+056 based on the equations discussed in earlier papers in the context of other blazars \citep[e.g.,][]{Abraham_2000, Caproni_Abraham_2004, Britzen_et_al_2018}, establishing a precession model for TXS\,0506+056 where a mildly relativistic jet ($\gamma\sim2.6$) crosses the line of sight every $\sim 5$ years (precession period of about 10 years).

Regarding  PKS\,0502+049, the right panel in \autoref{fig:TXS_PKS_RA_Dec_Distribution_figure} shows a narrower jet-component distribution along the right ascension direction for a fixed declination offset in comparison with that seen in TXS\,0506+056. Despite the geometrical interpretation (changes in the jet direction) mentioned previously, the wider scatter observed for the components of TXS\,0506+056 on the plane of the sky could also be a consequence of the different observational characteristics of these two blazars: as TXS\,0506+056 and PKS\,0502+049 are located at distinct redshifts, their angular scales are not the same (roughly a factor of two larger in the case of TXS\,0506+056), as well as the interval between first and last epoch of the maps used in this work (the observations of TXS\,0506+056 span a time range measured at its rest frame that is about a factor of 5 larger than the corresponding interval for PKS\,0502+049). In addition, the initial NE-SW orientation of the parsec-scale jet of PKS\,0502+049 suffers a clear and systematic bending towards south direction after a core distance of $\sim 2.2$ mas. 


\subsection{General constraints for the parsec-scale jets of TXS 0506+056 and PKS 0502+049}\label{sec:GenConstr}

All jet components here identified exhibit superluminal speeds ranging from $9.5c$ to $66.5c$ and from $14.3c$ to $59.1c$ for TXS\,0506+056 and PKS\,0502+049, respectively. 
The apparent superluminal velocities $\beta_{\mathrm{app}}$ depend on the Lorentz factor $\gamma$ and on the viewing angle $\theta$ through the relation:
\begin{equation}
	\beta_\mathrm{app} = \frac{\sin\theta}{\gamma\left(\gamma^2-1\right)^{-1/2}-\cos\theta} \; .
	\label{eq:apparent_speed}
\end{equation}
\noindent
where
\begin{equation}
	\gamma = {(1-\beta^2)^{-1/2}}
	\label{eq:gamma}
\end{equation}
\noindent
and $\beta$ is the bulk velocity of the jet.

The apparent velocities has a maximum value ${\beta_{\mathrm{app}}^\mathrm{max}}$ when $\cos\theta = \beta$; the minimum value of the relativistic Lorentz factor that satisfies this relation is: 
\begin{equation}
    \gamma_{\mathrm{min}} = \sqrt{1 + (\beta_{\mathrm{app}}^{\mathrm{max}})^2} \; ,
\end{equation}

Our fastest (and reliable) identified jet components, C11 in the case of TXS 0506+056 and C7 in PKS 0502+049, imply $\gamma_{\mathrm{min}} = 66 \pm 14$ and $\gamma_{\mathrm{min}} = 59 \pm 11$, respectively.
Note that the Lorentz factor estimated for TXS\,0506+056 differs from those found in \cite{Kun_et_al_2019} and \cite{Li_et_al_2020} ($\gamma \sim 5$), since the apparent speeds found in this work are substantially higher than those inferred in former studies. Potential reasons for such discrepancy might be related to the identification scheme for jet components adopted in this work (see \autoref{sec:kin}), as well as the usage of an image-based decomposition of the interferometric maps through the CE technique instead of visibility model-fits done in previous works. A denser monitoring of the parsec-scale jet of TXS\,0506+056 in future (similar to that seen in \autoref{fig:TXS_PKS_fitting_figure} after mid 2019, as well as perfomed for other sources; e.g., \citealt{2013A&A...559A..75B, 2017ApJ...846...98J}) is crucial for confirming the proposed jet-component identification in this work.

On the other hand, the maximum value of the angle between the jet orientation and the line of sight can be calculated from

\begin{equation}
\theta_{\mathrm{max}} \approx \arccos{\left[\frac{(\beta_{\mathrm{app}}^{\mathrm{min}})^{2} - 1}{(\beta_{\mathrm{app}}^{\mathrm{min}})^{2} + 1}\right]} \; ,
\label{eq:theta_max}
\end{equation} 
\\where $\beta_{\mathrm{app}}^{\mathrm{min}}$ is the lowest apparent speed among detected jet components.

We found  maximum viewing angles of $12\degr\pm 4\degr$ and $8\degr\pm 5\degr$ for TXS\,0506+056 and PKS\,0502+049, respectively. 
In the case of TXS\,0506+056, \cite{Li_et_al_2020} obtained a jet viewing angle $\theta_{\mathrm{max}}= 20\degr \pm 2\degr$, whereas \cite{Kun_et_al_2019} found $\theta_{\mathrm{max}} = 8\fdg2$ using the average Doppler factor of the core. 

Another relevant parameter related to jet kinematics is the Doppler boosting factor, $\delta$, defined as

\begin{equation}%
	\delta = \left[\gamma-\left(\gamma^2-1\right)^{1/2}\cos\theta\right]^{-1} \; .
	\label{eq:doppler_factor}
\end{equation}

Thus, estimates for $\delta$ associated with the robust jet components can be obtained from the previous knowledge of their jet viewing angles and Lorentz factors. Imposing $\gamma=\gamma_\mathrm{min}$ and using the values of $\beta_\mathrm{app}$ listed in \autoref{table:Kin_Param_TXS_PKS_table}, values of $\theta$ for each jet component can be determined from \autoref{eq:apparent_speed}.


Except for $\beta_\mathrm{app}=\beta_\mathrm{app}^\mathrm{max}=\sqrt{\gamma_\mathrm{min}^2-1}$, this procedure leads to two independent solutions for $\theta$, which translates consequently to two possible values for $\delta$ (see \autoref{eq:doppler_factor}). Considering all jet components found in this work, we derived possible ranges for $\delta$ in both sources, resulting to $0.5\la\delta\la 132$ for TXS\,0506+056 and $1.8\la\delta\la116$ for PKS\,0502+049. Note that values of $\gamma$ higher than $\gamma_\mathrm{min}$ would broaden those ranges since the upper (lower) limit for $\delta$ would be shifted upwards (downwards) in relation to those calculated above.

\subsection{Brightness temperature of the core region}\label{sec:BrightTemp}

The brightness temperature of the core region in the rest frame of the source, $T_\mathrm{B, rest}$, is defined as

\begin{equation}
	T_\mathrm{B, rest} = (1 + z) \, \frac{2 \, \ln{2}}{\pi k} \frac{c^2}{\nu^2} \frac{F}{a_{\mathrm{FWHM}} \, b_{\mathrm{FWHM}}} \; ,
	\label{eq:brightness_temperature_equation}
\end{equation}
\\where $z$ is the redshift of the host galaxy, $k$ is the Boltzmann constant, and $a_{\mathrm{FWHM}}$ and $b_{\mathrm{FWHM}}$ are, respectively, the FWHM of the elliptical Gaussian components along the major and the minor axes. In the case of an unresolved emitting region, the term $a_{\mathrm{FWHM}} \, b_{\mathrm{FWHM}}$ must be replaced by $d_{\mathrm{min}}^2$, which can be written as \citep{Lobanov_2005}

\begin{equation}
	d_{\mathrm{min}} = \frac{2^{2-\varpi/2}}{\pi} \left[ \pi \ln{2} \; \left(\Theta^\mathrm{FWHM}_{\mathrm{beam}}\right)^2 \sqrt{1 - \epsilon^2_{\mathrm{beam}}} \, \ln{\left( \frac{\mathrm{SNR}}{\mathrm{SNR} - 1} \right)} \right] ^{1/2} \;,
	\label{eq:dmin}
\end{equation}
\\where SNR (= $I_{\mathrm{max}}$/RMS) is the signal-to-noise ratio and $\varpi$ is an index describing the weighting method used to generate the radio maps ($\varpi=0$ for uniform weighting and $\varpi=2$ for natural weighting; \citealt{Lobanov_2005}).

We listed in \autoref{table:Parameters_PKS_8GHz_table} the values of $T_\mathrm{B, rest}$ for the core, calculated from \autoref{eq:brightness_temperature_equation}\footnote{Values preceded by the symbol ``\,$>$\,'' in Tables \ref{table:Parameters_TXS_table}, \ref{table:Parameters_PKS_15GHz_table} and \ref{table:Parameters_PKS_8GHz_table} indicate lower limit for the brightness temperature, which was calculated using $a^{\mathrm{FWHM}}_{\mathrm{beam}}b^{\mathrm{FWHM}}_{\mathrm{beam}}=d_{\mathrm{min}}^2$ in \autoref{eq:brightness_temperature_equation} when $a_{\mathrm{FWHM}} < d_{\mathrm{min}}$.}. It ranges from $0.6 \leq T_\mathrm{B,rest} (10^{11}\,$K$) \leq 8.5$ for TXS\,0506+056 and $2.8 \leq T_\mathrm{B,rest} (10^{11}\,$K$) \leq 39.8$ for PKS\,0502+049. 

\begin{figure*}
	\centering
	\includegraphics[width=1\textwidth]{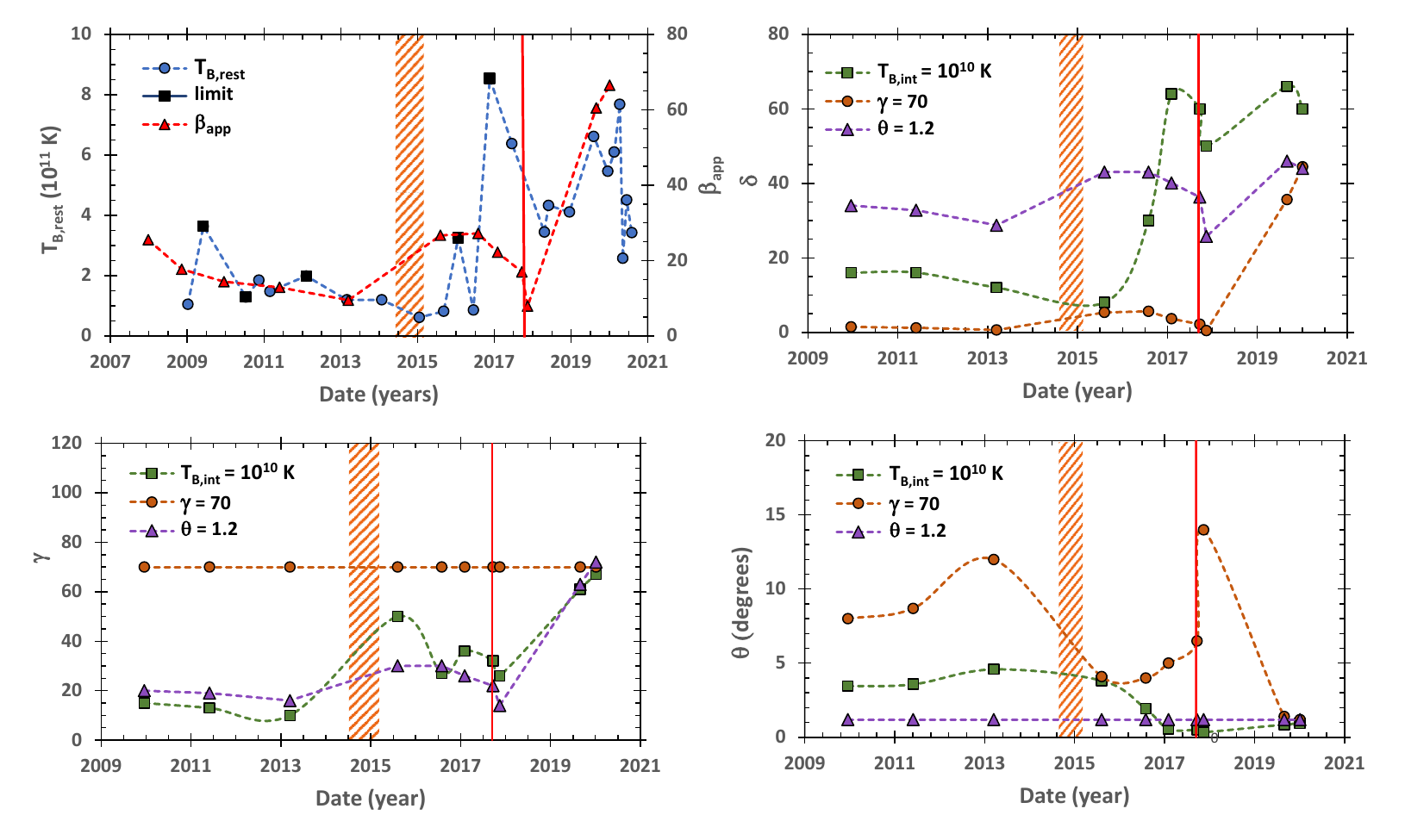}
    \caption{{\it Upper left panel:} The blue circles show the core's brightness temperature at rest frame of TXS\,0506+056 as a function of time. When core is unresolved, a lower limit for the brightness temperature is provided and marked by black squares. The apparent speed of each jet component of TXS\,0506+056 in terms of their respective ejection epochs are displayed by red triangles. {\it Upper right panel:} Doppler boosting factors derived at the ejections epochs of the jet components assuming a constant value for $T_\mathrm{B,int} (=10^{10})$ K (green squares), a constant value for $\gamma (=70)$ (red circles), and a constant value for $\theta (=1\fdg2)$ (purple triangles). {\it Lower left panel:} The same of the previous panel but showing the behaviour of the jet bulk Lorentz factor. {\it Lower right panel:} The same of the previous panel but showing the behaviour of the jet viewing angle. The epochs of the neutrino excess and the IceCube-170922A event are marked respectively by the hatched red rectangle and the vertical red line in all the panels.
    }
    \label{fig:TXS_Doppler_Gamma_Theta}
\end{figure*}

In the upper left panel of \autoref{fig:TXS_Doppler_Gamma_Theta}, we show the time behaviour of $T_\mathrm{B,rest}$, as well as $\beta_\mathrm{app}$ as a function of $t_0$ for TXS\,0506+056. Since the epochs of the  VLBI observations do not coincide with the ejection epochs of the superluminal components, the brightness temperature values are interpolations.   A clear increase of $T_\mathrm{B,rest}$ by a factor of 3  is seen after the year 2016, which also coincides with the IceCube-170922A event. A more complicated behaviour is noted in the case of $\beta_\mathrm{app}$, which presents non-monotonic variations during the whole period of the interferometric monitoring. 
We do not discuss the behaviour of PKS\,0502+049 because of the limited time coverage of the VLBI observations.

The measured brightness temperature is related to the intrinsic brightness temperature of the source at its rest frame $T_\mathrm{B,int}$ by the relation \citep[e.g.,][]{Readhead_1994, Kovalev_et_al_2005, Homan_et_al_2006}:


\begin{equation}
    T_\mathrm{B, rest} = \delta \, T_\mathrm{B,int} \; ,
    \label{eq:TBrest_delta_TBint_equation}
\end{equation}

\citet{Readhead_1994} studied the distribution of intrinsic brightness temperatures of sample of blazars and concluded that  they are concentrated around the equipartition value of $\sim 5\times 10^{10}$\,K, and proposed the use of \autoref{eq:TBrest_delta_TBint_equation} as a tool to calculate the Doppler factor. In the next subsection we apply this technique to calculate the variation of the physical parameters of the superluminal components of  TXS\,0506+056.

\begin{figure*}
	\centering
	\includegraphics[width=1\textwidth]{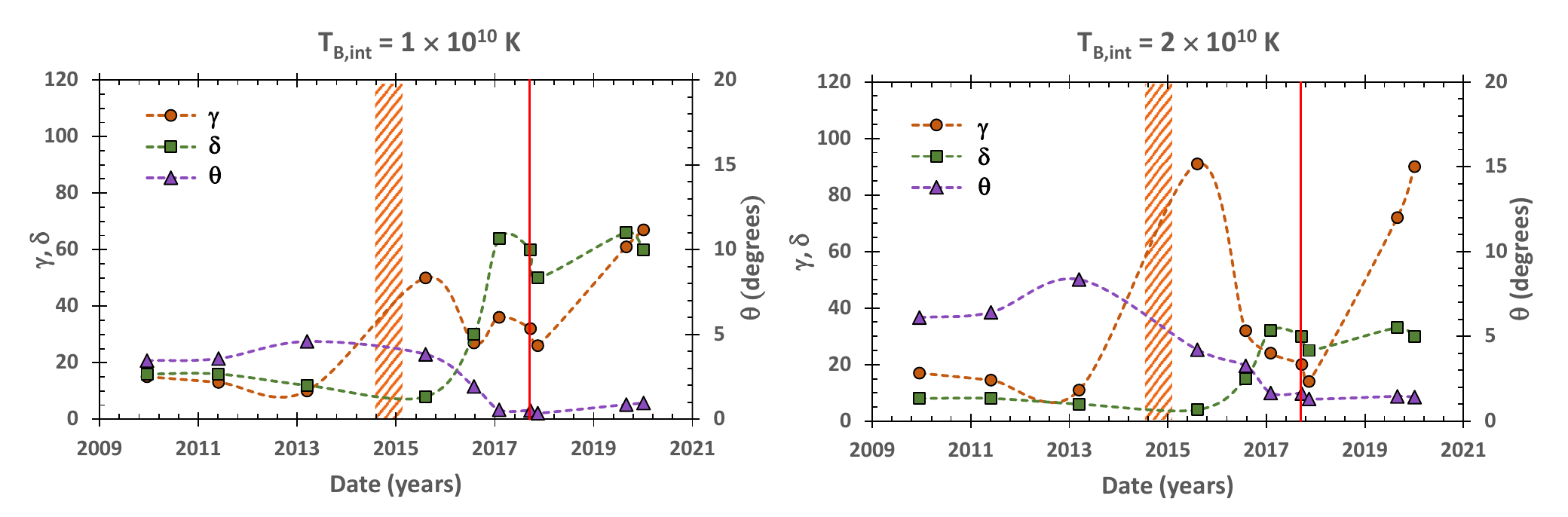}
    \caption{The values of Lorentz factor (red circles), Doppler factor (green squares), and jet viewing angle (purple triangles) at the ejection epochs of the jet components in TXS\,0506+056 considering $T_\mathrm{B,int} = 1\times 10^{10}$ K (left panel) and $T_\mathrm{B,int} = 2\times 10^{10}$ K (right panel). The epochs of the neutrino excess and the IceCube-170922A event are marked respectively by the hatched red rectangle and the vertical red line in both panels.
    }
    \label{fig:TXS_Doppler_Gamma_Theta_TB}
\end{figure*}

\subsection{Temporal changes of the jet parameters}
\label{sec:NewApprJetParam}

In blazars, the Doppler factor of the jet  is one of the most important input parameters for the models that aim to reproduce the observed neutrino observations  and their relation to different electromagnetic frequencies;
its value depends on  the Lorentz factor and  on  the jet direction relative to the line of sight. The same dependence affects the value of the apparent superluminal velocities.
The jet direction is assumed to be perpendicular to the accretion disk (e.g., \citealt{2000ApJ...534..165J}) and frequently the differences between the apparent superluminal velocities of the  jet components are attributed to differences in the viewing angles of the jet at the epoch at which the components were ejected (e.g., \citealt{2003ApJ...589L...9H}).  For many objects, the variations of the jet direction were attributed to precession; the Lorentz factor was assumed to be constant along the precessing period and its value was determined fitting the models (viewing and position angles on the plane of the sky)  to the observations; variations of the Doppler factor with time were calculated from the precessing model and the constant value of the Lorentz factor (e.g., \citealt{1982ApJ...262..478G, 2000A&A...360...57R, 2004ApJ...602..625C, 2011A&A...526A..51K, 2013A&A...557A..85R, Caproni_et_al_2017, Britzen_et_al_2018}).

In the case of TXS\,0506+056, it is not clear that the variations of $	\beta_\mathrm{app}$ with time are due to precession. In fact, even if the aperture of the precession cone is small, as implied from the minimum and maximum viewing angles, their  small absolute values require variations of the projected angle on the plane of the sky much larger than what is observed. 
However, we can still calculate the Doppler factor from \autoref{eq:TBrest_delta_TBint_equation}, assuming a constant intrinsic  brightness temperature for the blazar. We show the results of this calculation in the top right panel of \autoref{fig:TXS_Doppler_Gamma_Theta} (green squares), which shows the variation of the Doppler factor with time assuming  the constant value  $T_\mathrm{B, int}= 10^{10}$\,K. Once the Doppler factor is known, the Lorentz factor and the viewing angle can be calculated from Equations \ref{eq:doppler_factor} and \ref{eq:apparent_speed}, respectively. Their time dependence is shown as green squares at the bottom left and right panels of \autoref{fig:TXS_Doppler_Gamma_Theta}, respectively. 
The same results are shown in the left panel of \autoref{fig:TXS_Doppler_Gamma_Theta_TB}, where we  can also see in the right panel, the variation with time  of the Doppler and Lorentz factors, and the jet viewing angle, obtained using $2 \times 10^{10}$\,K for the constant intrinsic brightness temperature.
The  value of $10^{10}$\,K is the lowest compatible with the observations, since it renders a maximum value of the Lorentz factor $\gamma = 67$, close to what is necessary  to obtain  the observed maximum apparent velocity.
 As expected, the increase of $T_\mathrm{B,int}$ produces a decrease in $\delta$ (see \autoref{eq:TBrest_delta_TBint_equation}) for all epochs, leading to a systematic increase of $\theta$. The solution with the highest brightness temperature also predicts too extreme values for $\gamma$ ($\sim 90$) in two epochs, 2015.6 and 2020.0, arguing in favour of constant-$T_\mathrm{B,int}$ models with intrinsic brightness temperatures lower than $2\times10^{10}$ K.

The Doppler and Lorentz factors, as well as the viewing angle of the jet can also be calculated from Equations \ref{eq:apparent_speed}, \ref{eq:gamma} and \ref{eq:doppler_factor}, assuming a constant value for one of these parameters. This has been done and it is shown in the top right and bottom left and right panels of \autoref{fig:TXS_Doppler_Gamma_Theta} as red circles for constant $\gamma = 70$, and purple triangles for constant $\theta = 1.2\degr$. Again, the constant parameters  $\gamma$ and $\delta$ have values close to the minimum compatible with the observed maximum apparent velocity. 

Analysing the results for the Doppler factor presented in the top right panel of \autoref{fig:TXS_Doppler_Gamma_Theta}, we can see  that when the Lorentz factor is maintained constant, the value of $\delta$ remains low  $(< 10)$, except for the two last epochs in which it increases by a factor of 4; when it is the viewing angle that remains constant, $\delta$ has a large value during all the observing period, but when it is the intrinsic brightness temperature that remains constant, $\delta$ has a sharp increase between 2016 and 2017, which can explain, by boosting, the high emission in $\gamma$ rays that was observed starting at this epoch and which coincides with the neutrino detection.

Regarding the Lorentz factor, when the intrinsic brightness temperature is maintained constant, we can see an increase in $\gamma$ by a factor of three between 2013 and 2016, which maybe could be related to the neutrinos detected between 2014 and 2015. Both $T_\mathrm{B,int}$ and $\theta$ constant require a large increase in the Lorentz factor at latter epochs to account for the high value of the superluminal velocities.

Finally, when we compare the behaviour of the viewing angle when $T_\mathrm{B,int}$ or $\gamma$ are constant, we can see that when $T_\mathrm{B,int}$ is constant, $\theta$ remains almost constant at $4\degr$ until 2016 and then it  drops to about $1\degr$, while when $\gamma$ is maintained constant, $\theta$ reaches much larger values and have irregular variability.

It is important to note that the whole analyses above assumed that two out of the three free parameters used to describe $T_\mathrm{B,int}$ and $\beta_\mathrm{app}$ at the ejection epochs of the jet components in TXS\,0506+056 can change in time, which does not exclude the possibilities of just one or even all the three parameters could be time variable.

\section{Discussion}\label{sec:Disc}


As mentioned previously, our results regarding the kinematics of the parsec-scale jets of TXS\,0506+056 and PKS\,0502+049 show that ejections of new components coincide (at 1$\sigma$-level) with the appearance of flares in their respective gamma-ray light curves. Moreover, the formation epochs of the jet components C9 and C10 in TXS\,0506+056 are compatible with the occurrence of the IceCube-170922A event, suggesting a possible connection between the radio core activity and the production of the detected 290-TeV neutrino. A systematic and strong increase of the flux density of the parsec-scale core of TXS\,0506+056 seen after 2017 (see \autoref{fig:TXS_PKS_flux_figure}) also argues in favour of this possibility. 

Indeed, AGN core regions as sites of production of neutrinos have been recently proposed in the literature. \citet{2020ApJ...894..101P} analysed statistically the origin of IceCube neutrinos with energies higher than $\sim$200 TeV using VLBI data and single-dish radio observations of 18 AGNs.
They found that these AGNs usually have more intense parsec-scale cores in comparison with other radio-loud sources, 
with their sites of high-energy neutrinos residing at distances lower or of the order of some parsecs from the SMBH.
\citet{2021ApJ...908..157P} extended those analyses, including all neutrinos detected by IceCube collaboration between 2008 and 2015, as well as relaxing the lower limit of $\sim$200 TeV adopted previously. They found that the association between neutrino production and blazar activity also exists for lower neutrino energies as well. 

\citet{2020arXiv201204425N} proposed a scenario to explain the correlation between radio emission and high-energy neutrinos associated with AGNs reported by \citet{2020ApJ...894..101P}. Their model considers that both synchrotron emitting electrons and neutrinos originate from decays of charged pions produced in proton-proton interactions in parsec-scale relativistic jets. The exact location of the site where neutrinos are produced depends on the density profile of the interstellar medium of these galaxies, not exceeding parsec-scale distances from the SMBH. Previously, \citet{2018ApJ...864...84K} considered single-zone theoretical models (neutrinos and gamma-rays produced at the same site) to study the relationship between the IceCube-170922A event and the contemporaneous electromagnetic emission from near-infrared to gamma-ray energies. They found that a hybrid leptonic scenario (gamma-rays produced by inverse-Compton scatterings and high-energy neutrinos via a radiatively subdominant hadronic component) can fairly describe the observed electromagnetic and neutrino fluxes. \citet{2020arXiv200910523H} found that a fraction of neutrino events are not associated with strong radio flaring blazars after correlating the radio data from the Owens Valley Radio Observatory and Mets\"ahovi Radio Observatory blazar monitoring programs with the IceCube neutrino events. However, a random coincidence is strongly disfavour when a large amplitude radio flares in a blazar is observed together with a concomitant and spatially coincident neutrino event according to these authors.

Considering the results shown in \autoref{fig:TXS_Doppler_Gamma_Theta} and \autoref{fig:TXS_Doppler_Gamma_Theta_TB}, the bulk Lorentz factor of the parsec-scale jet of TXS\,0506+056 ranges roughly between 10 and 70, which agrees with values invoked in some theoretical models used to interpret the IceCube-170922A event \citep[e.g.,][]{2019PhRvD..99j3006B, 2020ApJ...889..118Z}. Regarding the Doppler boosting factor, those theoretical models have adopted values between $\sim5$ and 30, in good agreement with the cases $\theta=1\fdg2$ (\autoref{fig:TXS_Doppler_Gamma_Theta}) and $T_\mathrm{B,int}=2\times10^{10}$ K (\autoref{fig:TXS_Doppler_Gamma_Theta_TB}).

On the other hand, the neutrino excess detected by IceCube collaboration between September 2014 to March 2015 coincides with a relative quiescent phase seen in the gamma-ray light curve of TXS\,0506+056 \citep[e.g.,][]{Padovani_et_al_2018, 2019ApJ...880..103G}, as well as the lack of any substantial parsec-scale core activity and/or emergence of any new jet component (see \autoref{fig:TXS_PKS_flux_figure}, for instance). Indeed, neutrino production without a clear electromagnetic counterpart (an ``orphan" neutrino flare; e.g., \citealt{2020arXiv201103681X}) brought challenges to some theoretical (single-zone) models since an electromagnetic cascade that follows neutrino emission is expected in such situation \citep[e.g.,][]{IceCube_Collaboration_et_al_2018, Rodrigues_et_al_2019, 2019ApJ...881...46R, 2020arXiv201103681X}. 

Several theoretical models focusing on the 2014-2015 neutrino excess are available in the literature. For instance, 
\citet{2019ApJ...881...46R} 
concluded that the neutrino excess was probably generated through hadronic interactions of relativistic protons with a stationary (in relation to the rest frame of the galaxy) soft X-ray photon field.These authors also found that the predicted GeV gamma-ray flux produced by inverse Compton during pair cascades is too low in comparison to the observed Fermi-LAT flux from TXS\,0506+056, concluding that neutrinos and gamma-rays are probably generated by distinct physical processes.
Similarly, \citet{Rodrigues_et_al_2019} analysed the viability of leptohadronic models with three different geometries (one zone, compact-core, and additional external photon field) to explain simultaneously the spectral energy distribution of TXS\,0506+056 and the IceCube neutrino spectrum, as well as the quantity of neutrinos ($13\pm5$) produced in such event. Difficulties in fulfilling simultaneously the observation constraints from electromagnetic and neutrino spectra were found independently of the adopted model. Moreover, they concluded that obtaining more than two to five neutrino events during the period between September 2014 and March 2015 implies violating the multi-wavelength constraints. Similar difficulties in describing simultaneously electromagnetic data and neutrino emission produced at the same site have been faced by other recent works \citep[e.g.,][]{2020ApJ...891..115P}. In contrast, \citet{2020ApJ...889..118Z} proposed a model that provides an explanation for the 2014–2015 neutrino excess without disrespecting the X-ray and gamma-ray observational constraints. In their model, protons and other heavier nuclei are accelerated in a compact region (blob) of blazar jets. Their interaction with radiation fields are able to generate not only high-energy photons, but also electron–positron pairs, neutrinos, and neutrons. Neutrinos and neutrons can freely escape from the blob whereas the high-energy photons may be trapped by local opacity effects. 

One of the radio maps of TXS\,0506+056 analysed in this work was obtained on 2015 January 18 (2015.049), roughly in the middle of the reported neutrino-excess time window \citep{IceCube_Collaboration_2018}. Our CE modelling of this image revealed the presence of a superluminal jet component, C5, at a core-component angular distance of $\sim0.84$ mas, corresponding to a projected distance, $L_\mathrm{proj}$, of about 4.0 pc at the redshift of TXS\,0506+056. The deprojected core-component distance, $L$ (calculated from $L=L_\mathrm{proj}/\sin\theta$), has a lower limit of about 19 pc\footnote{The distance between jet inlet region (core) and the SMBH is typically smaller than $\sim10$ pc \citep[e.g.,][]{2012A&A...545A.113P, 2019MNRAS.485.1822P}, so that the distance between the SMBH and jet component C5 is probably a bit larger than this minimum value.} considering a jet viewing angle of $\sim12\degr$ (upper limit value derived from \autoref{eq:theta_max}), and an upper limit of about 191 pc considering the model with $\theta=1\fdg2$ shown in \autoref{fig:TXS_Doppler_Gamma_Theta}. The predicted Doppler factor during the neutrino excess ranges roughly from 5 to 40, depending of the case displayed in Figures \ref{fig:TXS_Doppler_Gamma_Theta} and \ref{fig:TXS_Doppler_Gamma_Theta_TB}. They agree with the adopted values in current neutrino models \citep[e.g.,][]{Rodrigues_et_al_2018, 2019ApJ...881...46R, 2020ApJ...889..118Z}. Even though the derived FWHM semi-major axis of C5 ($\sim0.38$ mas or $\sim 1.8$ pc) is substantially larger than usual radii of the blobs adopted in neutrino-production models \citep[$\la0.03$ pc; e.g.,][]{Murase_et_al_2018, 2019ApJ...881...46R, 2020ApJ...889..118Z}, we speculate that the parsec-scale jet component C5 in TXS\,0506+056 might be the site where 2014-2015 neutrino excess was produced. Of course this possibility must be validated by appropriated neutrino-production models in future works.

An alternative to TXS\,0506+056 as the unique driver of the 2014–2015 IceCube neutrino excess has also been proposed in the literature \citep[e.g.,][]{Liang_et_al_2018, He_et_al_2018, Banik_et_al_2020}. In this scenario, the gamma-ray loud PKS\,0502+049 (separated by $\sim1\fdg2$ from TXS\,0506+056) could be responsible for a fraction of the muon-neutrino events detected between September 2014 and March 2015. The median angular resolution of the muon-neutrino events detected by IceCube is about $0\fdg5$ at energies around $\sim30$ TeV \citep[e.g.,][]{IceCube_Collaboration_2018}. However, the directional reconstruction uncertainty for each individual neutrino event ranges from $\sim 0\fdg2$ to more than 2\degr \citep[e.g.,][]{Liang_et_al_2018}, which does not exclude the possibility that PKS\,0502+049 could have contributed to the observed neutrino excess. Indeed, \citet{Liang_et_al_2018} calculated the individual uncertainties regarding the neutrino trajectories and found that 7 out of the 13 neutrino events are spatially consistent with the sky position of PKS\,0502+049. This result agrees with \citet{Rodrigues_et_al_2019} who found difficulties in producing more than five neutrino events in their models without violating the multi-wavelength constraints for TXS\,0506+056 between the epochs 2014 and 2015. Interactions between the relativistic jet and dense clouds associated with the broad line region of PKS\,0502+049 have been invoked as drivers of neutrinos related to  2014-2015 Icecube event \citep{He_et_al_2018}, while \citet{Banik_et_al_2020} proposed a proton blazar jet scenario \citep[e.g.,][]{2001APh....15..121M, 2019PhRvD..99j3006B} where non relativistic protons would be targets for the production of high-energy gamma rays and neutrinos via proton-proton interactions (due to the presence of relativistic protons in the jet). \citet{Banik_et_al_2020} showed that neutrino excess observed during 2014–2015 is incompatible with the premise that TXS\,0506+056 would have been the unique source responsible for such neutrinos. Indeed, their results suggest that PKS\,0502+049 may be (partially or totally) responsible for the 2014–2015 neutrino excess. The adopted values $\delta=40$ and $\theta=1\fdg35$ by \citet{Banik_et_al_2020} are compatible with the values of these quantities derived in this present work ($2\lesssim\delta\lesssim116$, $\theta\lesssim8\degr$), even though their adopted value for the Lorentz factor ($\gamma=30$) is almost a factor of two lower than $\gamma_\mathrm{min}$ derived in this work ($\gamma\ga59$).

It is important to emphasise that the observed gamma-ray flaring state of PKS\,0502+049 has motivated several studies (as previously reported above) regarding the viability of this quasar as the main driver of the IceCube neutrino excess. In this work we report for the first time a possible connection between 2014-2015 gamma-ray flares and the formation of new jet components in PKS\,0502+049. We can note in \autoref{fig:TXS_PKS_flux_figure} that jet components C3 and C4 were ejected during the occurrence of gamma-ray flares, which could be the sites where electromagnetic/neutrino emissions take place in some neutrino production models. Unfortunately, there is no interferometric data of PKS\,0502+049 publicly available during the ejection period of C3 and C4 that allow us to infer any simultaneous flaring activity of the core region. 


\section{Conclusions}\label{sec:concl}

In this work, we presented the first kinematic study of the parsec-scale jet of the FSRQ PKS\,0502+049 based on 13 interferometric images at 8 and 15\,GHz. We have also thoroughly investigated the kinematic properties of the parsec-scale jet of TXS\,0506+056 at 15\,GHz, using 24 VLBA observations spanning the years 2009–2020. Based on the assumption that the jet emission can be modelled by discrete components described mathematically by two-dimensional elliptical Gaussian functions, we applied our CE global optimisation technique to determine the structural parameters of these Gaussian components.

The parsec-scale jet of PKS\,0502+049 is highly relativistic, exhibiting seven moving components with apparent speeds ranging from $14.3c$ and $59.1c$, and pointing close to the line of sight ($\theta \la$ 8\degr $\pm$ 5\degr). The maximum observed speed implies a minimum jet Lorentz factor of 59 $\pm$ 11 for this source.

In the case of TXS\,0506+056, we have identified twelve components in the parsec-scale jet receding ballistically from the core with superluminal apparent speeds (from $9.5c$ to $66c$). Through the fastest jet component we inferred a lower limit of 66 for the jet bulk Lorentz factor, and from the lowest apparent speed we obtained a conservative upper limit of $12\degr$ for the jet viewing angle, in agreement with previous estimates found in the literature. A novel approach using simultaneously the brightness temperature of the core region and the apparent speeds of the jet components allowed us to infer some basic jet parameters for TXS\,0506+056 at distinct epochs. These results are in full agreement with the limits of $\gamma$, $\theta$ and $\delta$ determined from jet kinematics only.
 
In addition, we analysed the Fermi-LAT data from 300 MeV to 300 GeV to generate gamma-ray light curves for TXS\,0506+056 and PKS\,0502+049. Our results showed that the occurrence of gamma-ray flares fall within $1\sigma$ uncertainties of the ejection epochs of new components in both blazars. Furthermore, the ejection of the jet components C9 and C10 in TXS\,0506+056 were contemporaneous with the occurrence of the IceCube-170922A event. Thus, we essentially suggest that the neutrino production may be connected with the radio core activity.

During the neutrino excess detected by the IceCube in 2014-2015, the blazar TXS\,0506+056 was found to be in a quiescent phase in gamma-rays. Neither parsec-scale core activity nor emergence of any new jet component were observed during this period. However, the presence of the jet component C5 at the arrival time window of such a neutrino excess might indicate it is a potential site of neutrinos. As already mentioned, this scenario must be validated by a proper neutrino-production model.

An alternative to TXS 0506+056 as the unique driver of such thirteen observed muon neutrinos has also been proposed in the literature \citep[e.g.,][]{Liang_et_al_2018, He_et_al_2018, Banik_et_al_2020}. As the nearby gamma-ray flaring blazar PKS\,0502+049 was in the state of enhanced emission at GeV energies during the reported neutrino excess time window, it could have contributed for this event too. Interestingly, we found that
this gamma-ray enhancement
was accompanied by ejections of two superluminal jet components (C3 and C4) in PKS\,0502+049, suggesting that these components could be jet blobs, namely the emitting region in some models of neutrino production in AGN.

In summary, although our findings strongly corroborate the previous association between TXS\,0506+056 and the IceCube-170922A event, they are still inconclusive regarding the 2014-2015 neutrino excess. We highlight that our results do not rule out the possibility of both sources, TXS\,0506+056 and PKS\,0502+049, had emitted these $13 \pm 5$ signal events quoted by the IceCube Collaboration as well.


\section*{Acknowledgements}

VYDS thanks Coordena\c{c}\~ao de Aperfei\c{c}oamento de Pessoal de N\'\i vel Superior (CAPES) for financial support. AC thanks grants \#2017/25651-5 and \#2014/11156-4, São Paulo Research Foundation (FAPESP), and ZA acknowledges Brazilian agencies FAPESP and CNPq. This research has made use of data from the MOJAVE database that is maintained by the MOJAVE team \citep[]{2018ApJS..234...12L}. The 8-GHz image of J0505+0415 publicly available at The Astrogeo VLBI FITS image database was generated by Leonid Petrov from a VLBA experiment on 2018 November 03. The authors thank the anonymous referee for his/her very constructive comments and suggestions that improved the presentation of this work.


\section*{Data Availability}

The data underlying this article were accessed from the MOJAVE
webpage (\url{http://www.physics.purdue.edu/MOJAVE/index.html}), the
Fermi data server (\url{https://fermi.gsfc.nasa.gov/cgi-bin/ssc/LAT/LATDataQuery.cgi}), and the Astrogeo Center (\url{http://astrogeo.org/vlbi_images/}). The data generated in this research will be shared on reasonable request to the corresponding author.




\bibliographystyle{mnras}
\bibliography{references} 




\appendix

\section{VLBI data}\label{appendix:VLBI}
We provide in this section the basic characteristics of the radio interferometric images of TXS\,0506+056 and PKS\,0502+049 analysed in this work, as well as the structural parameters of the elliptical Gaussian components found after CE optimisation of those images.

\begin{center}
    \onecolumn
        \small
		\begin{longtable}{ccccccc}
			\caption{Quantitative characteristics of the 24 radio images of the blazar TXS\,0506+056 analysed in this work.}
			\label{table:Characteristics_TXS_table} \\
			\toprule
			Epoch & Frequency & $\Theta^\mathrm{FWHM}_{\mathrm{beam}}$ & $\epsilon$ & $\theta$ & RMS & $I_{\mathrm{max}}$ \\
		      & (GHz) & (mas) &  & (deg) & (mJy beam$^{-1}$) & (Jy beam$^{-1}$) \\
			\midrule
			\endfirsthead
			\multicolumn{7}{c}%
			{\tablename\ \thetable\ -- \textit{(Continued)}} \\
			\toprule
        	Epoch & Frequency & $\Theta^\mathrm{FWHM}_{\mathrm{beam}}$ & $\epsilon$ & $\theta$ & RMS & $I_{\mathrm{max}}$ \\
		      & (GHz) & (mas) &  & (deg) & (mJy beam$^{-1}$) & (Jy beam$^{-1}$) \\
			\midrule
			\endhead
			\midrule \multicolumn{7}{r}{\textit{Continued on next page}} \\
			\endfoot
			\endlastfoot
		2009-01-07 & 15             & 1.40 & 0.89   & -3.87             & 0.15 & 0.42   \\
		2009-06-03 & 15             & 1.34 & 0.90   & -6.86             & 0.18 & 0.44   \\
		2010-07-12 & 15             & 1.38 & 0.93   & -8.48             & 0.18 & 0.27   \\
		2010-11-13 & 15             & 1.37 & 0.91   & -9.12             & 0.15 & 0.25   \\
		2011-02-27 & 15             & 1.18 & 0.89   & -0.76             & 0.21 & 0.23   \\
		2012-02-06 & 15             & 1.29 & 0.89   & -5.41             & 0.14 & 0.24   \\
		2013-02-28 & 15             & 1.24 & 0.90   & -3.80             & 0.16 & 0.24   \\
		2014-01-25 & 15             & 1.20 & 0.88   & -5.47             & 0.07 & 0.31   \\
		2015-01-18 & 15             & 1.34 & 0.92   & -4.78             & 0.08 & 0.30   \\
		2015-09-06 & 15             & 1.27 & 0.91   & -6.21             & 0.08 & 0.27   \\
		2016-01-22 & 15             & 1.33 & 0.91   & -9.78             & 0.07 & 0.21   \\
		2016-06-16 & 15             & 1.17 & 0.90   & -2.43             & 0.07 & 0.32   \\
		2016-11-18 & 15             & 1.19 & 0.90   & -4.74             & 0.07 & 0.40   \\
		2017-06-17 & 15             & 1.18 & 0.90   & -5.86             & 0.09 & 0.45   \\
		2018-04-22 & 15             & 1.21 & 0.90   & -5.64             & 0.09 & 0.71   \\
		2018-05-31 & 15             & 1.19 & 0.90   & -3.25             & 0.09 & 0.80   \\
		2018-12-16 & 15             & 1.21 & 0.88   & 2.44              & 0.13 & 0.84   \\
		2019-08-04 & 15             & 1.23 & 0.91   & -5.30             & 0.07 & 1.26   \\
		2019-12-17 & 15             & 1.33 & 0.89   & -6.40             & 0.08 & 1.59   \\
		2020-02-16 & 15             & 1.39 & 0.89   & -7.52             & 0.09 & 1.73   \\
		2020-04-09 & 15             & 1.78 & 0.90   & 22.82             & 0.09 & 1.85   \\
		2020-05-08 & 15             & 1.33 & 0.88   & -2.64             & 0.09 & 1.83   \\
		2020-06-13 & 15             & 1.23 & 0.92   & -2.60             & 0.10 & 1.49   \\
		2020-08-01 & 15             & 1.24 & 0.90   & 0.03              & 0.10 & 1.45   \\

		\bottomrule
		\end{longtable}
		
		\begin{longtable}{ccccccc}
			\caption{Quantitative characteristics of the 13 radio images of the blazar PKS\,0502+049 analysed in this work.} 	\label{table:Characteristics_PKS_table} \\
			\toprule
		  	Epoch & Frequency & $\Theta^\mathrm{FWHM}_{\mathrm{beam}}$ & $\epsilon$ & $\theta$ & RMS & $I_{\mathrm{max}}$ \\
		      & (GHz) & (mas) &  & (deg) & (mJy beam$^{-1}$) & (Jy beam$^{-1}$) \\
			\midrule
			\endfirsthead
			\multicolumn{7}{c}%
			{\tablename\ \thetable\ -- \textit{(Continued)}} \\
			\toprule
			Epoch & Frequency & $\Theta^\mathrm{FWHM}_{\mathrm{beam}}$ & $\epsilon$ & $\theta$ & RMS & $I_{\mathrm{max}}$ \\
		      & (GHz) & (mas) &  & (deg) & (mJy beam$^{-1}$) & (Jy beam$^{-1}$) \\
			\midrule
			\endhead
			\midrule \multicolumn{7}{r}{\textit{Continued on next page}} \\
			\endfoot
			\endlastfoot
		2016-09-26 & 15             & 1.15 & 0.90   & -5.47             & 0.09 & 0.66 \\
		2016-11-06 & 15             & 1.22 & 0.91   & -7.59             & 0.09 & 0.79 \\
		2016-12-10 & 15             & 1.44 & 0.89   & 12.14             & 0.09 & 0.81 \\
		2017-01-28 & 15             & 1.33 & 0.91   & -6.66             & 0.09 & 0.79 \\
		2018-04-22 & 15             & 1.19 & 0.90   & -4.89             & 0.09 & 0.78 \\
		2018-08-19 & 15             & 1.05 & 0.89   & -4.57             & 0.11 & 0.76 \\
		2018-11-03 & 8              & 2.00 & 0.90   & -2.16             & 0.46 & 0.50 \\
		2018-11-11 & 15             & 1.06 & 0.87   & 0.54              & 0.11 & 0.59 \\
		2019-04-15 & 15             & 1.19 & 0.89   & -2.55             & 0.11 & 0.53 \\
		2019-07-19 & 15             & 1.91 & 0.95   & -17.95            & 0.09 & 0.76 \\
		2020-05-25 & 15             & 1.96 & 0.94   & -18.84            & 0.09 & 0.70 \\
        2020-08-01 & 15             & 1.23 & 0.90   & 0.97              & 0.09 & 0.77 \\
        2020-01-12 & 15             & 1.24 & 0.89   & -5.67             & 0.07 & 0.58 \\
		\bottomrule
		\end{longtable}

	\end{center}

\begin{center}
    \onecolumn
        \small
		\begin{longtable}{ccccccccc}
			\caption{CE model-fitting jet parameters for the 15-GHz maps of TXS\,0506+056. Columns from left to right refer, respectively, to observation epoch, component label, flux density, component distance, position angle, FWHM major axis, axial ratio between minor and major axes, structural position angle and observed brightness temperature corrected to the rest frame of TXS\,0506+056.}
			\label{table:Parameters_TXS_table} \\
			\toprule
			Epoch & ID$^\mathrm{a}$ & $F$ & $r^\mathrm{b}$ & $\eta^\mathrm{c}$ & $a_\mathrm{FWHM}$ & Axial Ratio & $\mathrm{SPA}^\mathrm{c,d}$ & $T_\mathrm{B, rest}$ \\
			& & [Jy] & [mas] & [deg] & [mas] & & [deg] & $10^{11}\,\mathrm{K}$ \\
			\midrule
			\endfirsthead
			\multicolumn{9}{c}%
			{\tablename\ \thetable\ -- \textit{(Continued)}} \\
			\toprule
        	Epoch & ID$^\mathrm{a}$ & $F$ & $r^\mathrm{b}$ & $\eta^\mathrm{c}$ & $a_\mathrm{FWHM}$ & Axial Ratio & $\mathrm{SPA}^\mathrm{c,d}$ & $T_\mathrm{B, rest}$ \\
			& & [Jy] & [mas] & [deg] & [mas] & & [deg] & $10^{11}\,\mathrm{K}$ \\
			\midrule
			\endhead
			\midrule \multicolumn{9}{r}{\textit{Continued on next page}} \\
			\endfoot
			\endlastfoot
2009.019  & Core &     0.438  $\pm$    0.072     &      0.01  $\pm$     0.09     &     -66.6  $\pm$    906.9     &     0.174  $\pm$    0.005     &     1.000  $\pm$    0.005     &    -138.10  $\pm$      0.04      &     1.05  $\pm$     0.18  \\
 &   C1     &     0.064  $\pm$    0.009     &      1.29  $\pm$     0.10     &    -166.9  $\pm$      4.2     &     0.996  $\pm$    0.019     &     0.866  $\pm$    0.005     &    -150.7  $\pm$      4.5  & -\\
 &      U      &     0.032  $\pm$    0.004     &      2.63  $\pm$     0.11     &     177.2  $\pm$      2.3     &     1.281  $\pm$    0.075     &     1.000  $\pm$    0.002     &     -70.8  $\pm$      3.3  & -\\
 2009.422  & Core &     0.319  $\pm$    0.062     &      0.09  $\pm$     0.09     &      10.7  $\pm$     55.0     &     0.012  $\pm$    0.080     &     0.899  $\pm$    0.053     &      -7.2  $\pm$     21.0      &      >     3.64  \\
 &   C2     &     0.182  $\pm$    0.044     &      0.31  $\pm$     0.10     &    -166.7  $\pm$     16.9     &     0.330  $\pm$    0.014     &     1.000  $\pm$    0.001     &    -162.0  $\pm$     24.7  & -\\
 &   C1     &     0.046  $\pm$    0.008     &      1.64  $\pm$     0.09     &    -163.5  $\pm$      3.2     &     0.965  $\pm$    0.038     &     0.866  $\pm$    0.002     &     -17.8  $\pm$      3.9  & -\\
 &      U      &     0.041  $\pm$    0.015     &      2.57  $\pm$     0.10     &     175.7  $\pm$      2.1     &     1.587  $\pm$    0.042     &     0.870  $\pm$    0.041     &     -64.8  $\pm$     13.9  & -\\
 2010.529  & Core &     0.157  $\pm$    0.050     &      0.14  $\pm$     0.10     &       9.8  $\pm$     35.3     &     0.014  $\pm$    0.104     &     0.898  $\pm$    0.066     &     -10.3  $\pm$     22.3      &      >     1.29  \\
 &   C3     &     0.149  $\pm$    0.053     &      0.22  $\pm$     0.11     &    -167.8  $\pm$     22.8     &     0.247  $\pm$    0.012     &     1.000  $\pm$    0.001     &    -165.2  $\pm$     24.1  & -\\
 &   C2     &     0.089  $\pm$    0.016     &      1.29  $\pm$     0.08     &    -164.9  $\pm$      3.7     &     1.095  $\pm$    0.014     &     0.866  $\pm$    0.006     &     -17.3  $\pm$      2.2  & -\\
 &   C1     &     0.029  $\pm$    0.005     &      2.88  $\pm$     0.09     &     174.9  $\pm$      1.7     &     1.747  $\pm$    0.035     &     0.867  $\pm$    0.010     &     -37.8  $\pm$      5.8  & -\\
 2010.869  & Core &     0.229  $\pm$    0.040     &      0.04  $\pm$     0.09     &     159.0  $\pm$    142.8     &     0.101  $\pm$    0.007     &     0.878  $\pm$    0.035     &      -0.7  $\pm$      3.3      &     1.85  $\pm$     0.42  \\
 &   C3     &     0.051  $\pm$    0.009     &      0.64  $\pm$     0.09     &    -168.3  $\pm$      8.0     &     0.092  $\pm$    0.038     &     0.984  $\pm$    0.061     &     -56.0  $\pm$      0.3  & -\\
 &   C2     &     0.053  $\pm$    0.007     &      1.90  $\pm$     0.09     &    -172.9  $\pm$      2.7     &     0.890  $\pm$    0.026     &     0.913  $\pm$    0.069     &     -54.8  $\pm$     15.3  & -\\
 &      U      &     0.023  $\pm$    0.004     &      1.29  $\pm$     0.09     &    -147.6  $\pm$      4.0     &     0.351  $\pm$    0.033     &     0.870  $\pm$    0.046     &     -19.2  $\pm$     20.3  & -\\
 &   C1     &     0.020  $\pm$    0.002     &      3.60  $\pm$     0.10     &     168.9  $\pm$      1.6     &     1.896  $\pm$    0.094     &     0.998  $\pm$    0.012     &    -122.1  $\pm$      8.5  & -\\
 2011.159  & Core &     0.228  $\pm$    0.038     &      0.01  $\pm$     0.08     &     176.1  $\pm$    544.6     &     0.101  $\pm$    0.007     &     0.998  $\pm$    0.014     &    -135.0  $\pm$      2.1      &      >     1.48  \\
 &   C3     &     0.068  $\pm$    0.007     &      0.84  $\pm$     0.08     &    -162.7  $\pm$      5.4     &     0.725  $\pm$    0.024     &     1.000  $\pm$    0.000     &     -22.7  $\pm$     28.1  & -\\
 &   C2     &     0.045  $\pm$    0.005     &      1.99  $\pm$     0.09     &    -169.5  $\pm$      2.3     &     0.862  $\pm$    0.054     &     0.999  $\pm$    0.019     &     -55.6  $\pm$      6.1  & -\\
 &   C1     &     0.027  $\pm$    0.007     &      3.91  $\pm$     0.11     &     173.7  $\pm$      1.4     &     2.430  $\pm$    0.122     &     0.868  $\pm$    0.025     &    -163.2  $\pm$     11.4  & -\\
 2012.101  & Core &     0.231  $\pm$    0.040     &      0.01  $\pm$     0.09     &     162.1  $\pm$    421.0     &     0.038  $\pm$    0.031     &     0.891  $\pm$    0.164     &      -2.2  $\pm$     17.0      &      >     1.99  \\
 &   C4     &     0.029  $\pm$    0.006     &      0.69  $\pm$     0.10     &    -165.8  $\pm$      7.4     &     0.240  $\pm$    0.089     &     0.872  $\pm$    0.044     &     -56.9  $\pm$      6.4  & -\\
 &   C3     &     0.049  $\pm$    0.009     &      1.57  $\pm$     0.12     &    -165.7  $\pm$      3.3     &     0.949  $\pm$    0.054     &     0.866  $\pm$    0.002     &     -16.2  $\pm$      3.7  & -\\
 &   C2     &     0.030  $\pm$    0.007     &      2.66  $\pm$     0.19     &     179.9  $\pm$      2.7     &     1.366  $\pm$    0.061     &     0.867  $\pm$    0.021     &     -35.9  $\pm$      6.8  & -\\
 2013.162  & Core &     0.236  $\pm$    0.040     &      0.02  $\pm$     0.08     &       3.1  $\pm$    199.2     &     0.120  $\pm$    0.007     &     0.993  $\pm$    0.051     &    -138.3  $\pm$      2.8      &     1.20  $\pm$     0.25  \\
 &   C4     &     0.043  $\pm$    0.005     &      0.74  $\pm$     0.08     &    -163.0  $\pm$      6.5     &     0.645  $\pm$    0.024     &     1.000  $\pm$    0.001     &     -28.9  $\pm$     29.2  & -\\
 &   C3     &     0.040  $\pm$    0.009     &      2.14  $\pm$     0.08     &    -172.9  $\pm$      2.2     &     1.111  $\pm$    0.024     &     0.867  $\pm$    0.015     &     -60.6  $\pm$      3.3  & -\\
 &   C2     &     0.023  $\pm$    0.007     &      3.64  $\pm$     0.13     &     172.2  $\pm$      1.7     &     2.553  $\pm$    0.141     &     0.871  $\pm$    0.054     &    -154.2  $\pm$     15.3  & -\\
 2014.068  & Core &     0.278  $\pm$    0.050     &      0.05  $\pm$     0.08     &      13.2  $\pm$     88.2     &     0.139  $\pm$    0.021     &     0.867  $\pm$    0.005     &    -150.3  $\pm$      6.1      &     1.20  $\pm$     0.43  \\
 &   C5     &     0.074  $\pm$    0.024     &      0.35  $\pm$     0.12     &    -165.2  $\pm$     15.3     &     0.391  $\pm$    0.057     &     0.941  $\pm$    0.228     &    -144.9  $\pm$     28.7  & -\\
 &   C4     &     0.039  $\pm$    0.007     &      1.50  $\pm$     0.10     &    -166.0  $\pm$      3.2     &     1.154  $\pm$    0.045     &     0.866  $\pm$    0.005     &      -8.9  $\pm$      4.3  & -\\
 &   C3     &     0.029  $\pm$    0.004     &      2.77  $\pm$     0.11     &     176.4  $\pm$      2.2     &     1.783  $\pm$    0.080     &     1.000  $\pm$    0.002     &     -80.2  $\pm$      5.9  & -\\
 2015.049  & Core &     0.311  $\pm$    0.054     &      0.04  $\pm$     0.08     &      -2.8  $\pm$    121.9     &     0.191  $\pm$    0.007     &     0.999  $\pm$    0.009     &    -139.1  $\pm$      0.1      &     0.62  $\pm$     0.12  \\
 &   C5     &     0.062  $\pm$    0.007     &      0.84  $\pm$     0.10     &    -168.4  $\pm$      5.9     &     0.761  $\pm$    0.042     &     0.999  $\pm$    0.011     &    -133.6  $\pm$      1.9  & -\\
 &   C4     &     0.040  $\pm$    0.004     &      2.21  $\pm$     0.10     &    -171.1  $\pm$      2.3     &     1.298  $\pm$    0.066     &     1.000  $\pm$    0.004     &     -65.5  $\pm$      3.5  & -\\
 2015.682  & Core &     0.281  $\pm$    0.048     &      0.01  $\pm$     0.08     &     170.4  $\pm$    555.5     &     0.170  $\pm$    0.002     &     0.866  $\pm$    0.005     &    -177.2  $\pm$      2.7      &     0.82  $\pm$     0.14  \\
 &   C5     &     0.066  $\pm$    0.013     &      1.00  $\pm$     0.08     &    -172.0  $\pm$      4.7     &     0.989  $\pm$    0.024     &     0.867  $\pm$    0.016     &    -129.7  $\pm$      3.0  & -\\
 &   C4     &     0.035  $\pm$    0.007     &      2.59  $\pm$     0.11     &     179.8  $\pm$      2.1     &     1.627  $\pm$    0.144     &     0.867  $\pm$    0.038     &     -40.5  $\pm$      8.0  & -\\
 2016.060  & Core &     0.196  $\pm$    0.035     &      0.03  $\pm$     0.09     &    -154.0  $\pm$    184.2     &     0.040  $\pm$    0.019     &     0.900  $\pm$    0.195     &      -4.6  $\pm$     28.0      &      >     3.25  \\
 &   C6     &     0.040  $\pm$    0.007     &      0.58  $\pm$     0.09     &    -168.0  $\pm$      8.4     &     0.160  $\pm$    0.026     &     0.887  $\pm$    0.124     &     -58.6  $\pm$      0.3  & -\\
 &   C5     &     0.047  $\pm$    0.008     &      1.42  $\pm$     0.09     &    -165.3  $\pm$      3.4     &     0.973  $\pm$    0.012     &     0.866  $\pm$    0.005     &     -11.5  $\pm$      3.2  & -\\
 &   C4     &     0.027  $\pm$    0.005     &      3.13  $\pm$     0.09     &     177.3  $\pm$      1.6     &     1.860  $\pm$    0.042     &     0.867  $\pm$    0.016     &     -13.9  $\pm$      5.9  & -\\
 2016.459  & Core &     0.338  $\pm$    0.055     &      0.00  $\pm$     0.08     &     106.6  $\pm$   3123.9     &     0.181  $\pm$    0.005     &     0.867  $\pm$    0.019     &    -178.7  $\pm$      3.0      &     0.86  $\pm$     0.15  \\
 &   C6     &     0.082  $\pm$    0.015     &      1.29  $\pm$     0.08     &    -170.2  $\pm$      3.5     &     1.265  $\pm$    0.024     &     0.866  $\pm$    0.009     &    -170.5  $\pm$      3.3  & -\\
 2016.883  & Core &     0.249  $\pm$    0.081     &      0.12  $\pm$     0.09     &      -2.8  $\pm$     38.8     &     0.035  $\pm$    0.125     &     0.869  $\pm$    0.006     &      -1.1  $\pm$      3.6      &      >     8.54  \\
 &   C7     &     0.203  $\pm$    0.085     &      0.23  $\pm$     0.12     &     175.5  $\pm$     20.3     &     0.261  $\pm$    0.028     &     1.000  $\pm$    0.000     &    -166.4  $\pm$     26.4  & -\\
 &   C6     &     0.056  $\pm$    0.010     &      1.43  $\pm$     0.08     &    -167.0  $\pm$      3.3     &     1.217  $\pm$    0.031     &     0.866  $\pm$    0.001     &     -22.3  $\pm$      3.2  & -\\
 &   C4     &     0.025  $\pm$    0.005     &      3.37  $\pm$     0.12     &     174.7  $\pm$      1.4     &     1.969  $\pm$    0.035     &     0.867  $\pm$    0.006     &      -0.3  $\pm$      2.6  & -\\
 2017.460  & Core &     0.358  $\pm$    0.064     &      0.06  $\pm$     0.08     &      -7.3  $\pm$     78.1     &     0.068  $\pm$    0.016     &     0.873  $\pm$    0.031     &     -42.8  $\pm$     23.1      &     6.38  $\pm$     3.29  \\
 &   C8     &     0.145  $\pm$    0.029     &      0.36  $\pm$     0.08     &     178.5  $\pm$     12.4     &     0.268  $\pm$    0.045     &     0.943  $\pm$    0.212     &    -174.4  $\pm$     27.1  & -\\
 &   C7     &     0.060  $\pm$    0.010     &      1.18  $\pm$     0.08     &    -168.6  $\pm$      3.8     &     1.111  $\pm$    0.014     &     0.866  $\pm$    0.002     &     -28.5  $\pm$      3.2  & -\\
 &   C6     &     0.006  $\pm$    0.001     &      2.12  $\pm$     0.09     &    -167.5  $\pm$      2.2     &     0.012  $\pm$    0.082     &     0.960  $\pm$    0.103     &     -54.4  $\pm$      9.5  & -\\
 &   C4     &     0.025  $\pm$    0.005     &      3.52  $\pm$     0.09     &     175.2  $\pm$      1.4     &     2.070  $\pm$    0.061     &     0.867  $\pm$    0.017     &    -178.8  $\pm$      6.6  & -\\
 2018.307  & Core &     0.618  $\pm$    0.105     &      0.08  $\pm$     0.08     &      -1.7  $\pm$     56.4     &     0.122  $\pm$    0.002     &     0.867  $\pm$    0.011     &     -18.8  $\pm$      6.6      &     3.45  $\pm$     0.60  \\
 &   C9     &     0.158  $\pm$    0.028     &      0.38  $\pm$     0.08     &     177.4  $\pm$     11.9     &     0.153  $\pm$    0.019     &     0.884  $\pm$    0.152     &      -2.1  $\pm$     17.9  & -\\
 &   C8     &     0.095  $\pm$    0.015     &      1.21  $\pm$     0.08     &    -171.9  $\pm$      3.8     &     1.312  $\pm$    0.014     &     0.866  $\pm$    0.003     &     -40.5  $\pm$      3.5  & -\\
 &   C7     &     0.010  $\pm$    0.002     &      2.20  $\pm$     0.09     &    -167.3  $\pm$      2.2     &     0.012  $\pm$    0.068     &     0.965  $\pm$    0.096     &    -162.7  $\pm$     25.3  & -\\
 &   C6     &     0.019  $\pm$    0.004     &      3.64  $\pm$     0.10     &     175.6  $\pm$      1.5     &     1.710  $\pm$    0.106     &     0.871  $\pm$    0.053     &      -0.8  $\pm$     10.6  & -\\
 2018.414  & Core &     0.718  $\pm$    0.122     &      0.06  $\pm$     0.08     &      -0.8  $\pm$     71.2     &     0.118  $\pm$    0.002     &     0.868  $\pm$    0.017     &      -0.2  $\pm$      2.7      &     4.33  $\pm$     0.76  \\
 &   C9     &     0.146  $\pm$    0.024     &      0.41  $\pm$     0.08     &     177.8  $\pm$     10.9     &     0.205  $\pm$    0.009     &     0.869  $\pm$    0.039     &     -34.8  $\pm$     10.0  & -\\
 &   C8     &     0.092  $\pm$    0.008     &      1.30  $\pm$     0.08     &    -173.0  $\pm$      3.5     &     1.220  $\pm$    0.031     &     1.000  $\pm$    0.002     &    -174.1  $\pm$     15.2  & -\\
 &   C7     &     0.007  $\pm$    0.002     &      2.22  $\pm$     0.38     &    -167.7  $\pm$      4.1     &     0.009  $\pm$    0.153     &     0.965  $\pm$    0.104     &     -20.6  $\pm$     32.1  & -\\
 &   C6     &     0.019  $\pm$    0.004     &      3.40  $\pm$     0.13     &     174.5  $\pm$      1.7     &     1.787  $\pm$    0.115     &     0.868  $\pm$    0.026     &    -151.0  $\pm$     13.3  & -\\
 2018.959  & Core &     0.736  $\pm$    0.125     &      0.07  $\pm$     0.08     &      -1.5  $\pm$     70.1     &     0.122  $\pm$    0.007     &     0.868  $\pm$    0.009     &    -168.5  $\pm$      9.2      &     4.11  $\pm$     0.85  \\
 &  C10     &     0.166  $\pm$    0.043     &      0.35  $\pm$     0.11     &     177.7  $\pm$     13.7     &     0.226  $\pm$    0.016     &     0.876  $\pm$    0.097     &     -33.9  $\pm$     11.6  & -\\
 &   C9     &     0.069  $\pm$    0.012     &      1.08  $\pm$     0.09     &     175.9  $\pm$      4.5     &     0.911  $\pm$    0.080     &     0.997  $\pm$    0.021     &     -21.5  $\pm$     27.4  & -\\
 &   C8     &     0.020  $\pm$    0.003     &      2.10  $\pm$     0.09     &    -166.9  $\pm$      2.3     &     0.325  $\pm$    0.052     &     0.992  $\pm$    0.046     &    -131.4  $\pm$     16.1  & -\\
 &   C7     &     0.029  $\pm$    0.008     &      3.07  $\pm$     0.13     &     173.4  $\pm$      2.3     &     2.385  $\pm$    0.188     &     0.999  $\pm$    0.005     &     -93.6  $\pm$     14.8  & -\\
 2019.592  & Core &     1.278  $\pm$    0.219     &      0.02  $\pm$     0.08     &     171.7  $\pm$    214.9     &     0.127  $\pm$    0.002     &     0.866  $\pm$    0.003     &    -147.8  $\pm$      4.5      &     6.62  $\pm$     1.16  \\
 &  C10     &     0.064  $\pm$    0.012     &      0.68  $\pm$     0.09     &    -178.8  $\pm$      6.8     &     0.226  $\pm$    0.038     &     0.898  $\pm$    0.157     &     -57.1  $\pm$      1.8  & -\\
 &   C9     &     0.067  $\pm$    0.008     &      1.66  $\pm$     0.09     &    -173.5  $\pm$      2.8     &     0.954  $\pm$    0.054     &     1.000  $\pm$    0.001     &     -10.9  $\pm$     25.9  & -\\
 2019.962  & Core &     1.621  $\pm$    0.269     &      0.00  $\pm$     0.09     &    -166.2  $\pm$   1037.5     &     0.158  $\pm$    0.002     &     0.866  $\pm$    0.006     &    -177.5  $\pm$      2.6      &     5.46  $\pm$     0.92  \\
 &  C11     &     0.113  $\pm$    0.015     &      0.91  $\pm$     0.09     &     171.4  $\pm$      5.6     &     0.537  $\pm$    0.014     &     1.000  $\pm$    0.001     &    -170.6  $\pm$     19.9  & -\\
 &   C9     &     0.023  $\pm$    0.004     &      1.70  $\pm$     0.09     &    -163.9  $\pm$      3.0     &     0.005  $\pm$    0.042     &     0.971  $\pm$    0.093     &     -54.3  $\pm$     10.8  & -\\
 &   C8     &     0.031  $\pm$    0.006     &      2.63  $\pm$     0.12     &     172.9  $\pm$      2.2     &     1.834  $\pm$    0.113     &     0.867  $\pm$    0.018     &      -2.1  $\pm$      7.4  & -\\
 2020.128  & Core &     1.655  $\pm$    0.281     &      0.02  $\pm$     0.09     &     129.1  $\pm$    226.6     &     0.141  $\pm$    0.026     &     0.987  $\pm$    0.091     &      -1.1  $\pm$      3.4      &     6.10  $\pm$     2.53  \\
 &  C12     &     0.190  $\pm$    0.057     &      0.47  $\pm$     0.11     &    -173.5  $\pm$     11.3     &     0.071  $\pm$    0.068     &     0.934  $\pm$    0.123     &     -56.1  $\pm$      0.9  & -\\
 &  C11     &     0.113  $\pm$    0.020     &      1.36  $\pm$     0.10     &    -177.5  $\pm$      4.0     &     1.319  $\pm$    0.033     &     0.867  $\pm$    0.016     &    -173.2  $\pm$      5.1  & -\\
 2020.273  & Core &     1.709  $\pm$    0.296     &      0.01  $\pm$     0.12     &     -93.5  $\pm$    561.9     &     0.137  $\pm$    0.016     &     0.867  $\pm$    0.006     &    -179.6  $\pm$      2.8      &     7.68  $\pm$     2.28  \\
 &  C12     &     0.307  $\pm$    0.064     &      0.59  $\pm$     0.14     &     179.7  $\pm$     11.4     &     0.466  $\pm$    0.085     &     0.954  $\pm$    0.205     &    -173.3  $\pm$     35.9  & -\\
 &  C11     &     0.066  $\pm$    0.013     &      1.76  $\pm$     0.14     &    -177.8  $\pm$      3.9     &     1.321  $\pm$    0.078     &     0.867  $\pm$    0.017     &      -7.4  $\pm$      5.8  & -\\
 2020.352  & Core &     1.377  $\pm$    0.398     &      0.08  $\pm$     0.10     &       4.4  $\pm$     63.9     &     0.233  $\pm$    0.005     &     0.714  $\pm$    0.001     &    -176.3  $\pm$      3.3      &     2.58  $\pm$     0.75  \\
 &      U      &     0.592  $\pm$    0.337     &      0.22  $\pm$     0.16     &    -175.1  $\pm$     23.6     &     0.210  $\pm$    0.040     &     0.718  $\pm$    0.022     &     -47.9  $\pm$      1.6  & -\\
 &  C12     &     0.104  $\pm$    0.032     &      1.10  $\pm$     0.15     &     171.1  $\pm$      7.9     &     1.121  $\pm$    0.122     &     0.715  $\pm$    0.011     &     -22.2  $\pm$     12.6  & -\\
 &  C11     &     0.028  $\pm$    0.006     &      1.97  $\pm$     0.11     &    -168.8  $\pm$      2.7     &     0.019  $\pm$    0.115     &     0.903  $\pm$    0.221     &     -49.2  $\pm$      2.1  & -\\
 &   C8     &     0.015  $\pm$    0.004     &      3.70  $\pm$     0.11     &     170.8  $\pm$      1.5     &     1.111  $\pm$    0.226     &     0.765  $\pm$    0.243     &      -4.5  $\pm$     20.1  & -\\
 2020.451  & Core &     1.175  $\pm$    0.244     &      0.04  $\pm$     0.08     &      18.5  $\pm$    115.7     &     0.146  $\pm$    0.026     &     0.887  $\pm$    0.024     &    -137.0  $\pm$      0.9      &     4.51  $\pm$     1.86  \\
 &      U      &     0.608  $\pm$    0.185     &      0.39  $\pm$     0.11     &     173.2  $\pm$     11.7     &     0.393  $\pm$    0.033     &     0.872  $\pm$    0.022     &     -49.3  $\pm$      0.4  & -\\
 &  C12     &     0.076  $\pm$    0.025     &      1.63  $\pm$     0.14     &     179.9  $\pm$      3.4     &     1.114  $\pm$    0.080     &     0.910  $\pm$    0.052     &    -111.6  $\pm$      4.4  & -\\
 2020.585  & Core &     1.144  $\pm$    0.218     &      0.04  $\pm$     0.08     &      11.4  $\pm$    117.4     &     0.167  $\pm$    0.024     &     0.866  $\pm$    0.001     &    -142.5  $\pm$      4.5      &     3.43  $\pm$     1.17  \\
 &      U      &     0.607  $\pm$    0.155     &      0.41  $\pm$     0.11     &     176.2  $\pm$     11.6     &     0.417  $\pm$    0.033     &     1.0000  $\pm$    0.0001     &    -166.3  $\pm$     27.3  & -\\
 &  C12     &     0.071  $\pm$    0.015     &      1.68  $\pm$     0.16     &     178.9  $\pm$      3.2     &     0.975  $\pm$    0.068     &     1.000  $\pm$    0.001     &    -117.8  $\pm$      3.5  & -\\
 
		\bottomrule
		\end{longtable}

		\begin{longtable}{ccccccccc}
			\caption{CE model-fitting jet parameters for the 15-GHz maps of PKS\,0502+049. Columns from left to right refer, respectively, to observation epoch, component label, flux density, component distance, position angle, FWHM major axis, axial ratio between minor and major axes, structural position angle and observed brightness temperature corrected to the rest frame of PKS\,0502+049.} 	\label{table:Parameters_PKS_15GHz_table} \\
			\toprule
		    Epoch & ID$^\mathrm{a}$ & $F$ & $r^\mathrm{b}$ & $\eta^\mathrm{c}$ & $a_\mathrm{FWHM}$ & Axial Ratio & $\mathrm{SPA}^\mathrm{c,d}$ & $T_\mathrm{B, rest}$ \\
			& & [Jy] & [mas] & [deg] & [mas] & & [deg] & $10^{11}\,\mathrm{K}$ \\
			\midrule
			\endfirsthead
			\multicolumn{9}{c}%
			{\tablename\ \thetable\ -- \textit{(Continued)}} \\
			\toprule
			Epoch & ID$^\mathrm{a}$ & $F$ & $r^\mathrm{b}$ & $\eta^\mathrm{c}$ & $a_\mathrm{FWHM}$ & Axial Ratio & $\mathrm{SPA}^\mathrm{c,d}$ & $T_\mathrm{B, rest}$ \\
			& & [Jy] & [mas] & [deg] & [mas] & & [deg] & $10^{11}\,\mathrm{K}$ \\
			\midrule
			\endhead
			\midrule \multicolumn{9}{r}{\textit{Continued on next page}} \\
			\endfoot
			\endlastfoot
 2016.738  & Core &     0.634  $\pm$    0.109     &      0.02  $\pm$     0.08     &      45.0  $\pm$    213.9     &     0.080  $\pm$    0.009     &     0.939  $\pm$    0.178     &      -2.7  $\pm$     18.9      &    11.18  $\pm$     3.89  \\
 &   C3     &     0.141  $\pm$    0.023     &      0.43  $\pm$     0.08     &    -135.4  $\pm$     10.0     &     0.191  $\pm$    0.021     &     0.996  $\pm$    0.040     &    -150.8  $\pm$     22.1  & -\\
 &   C2     &     0.028  $\pm$    0.004     &      2.33  $\pm$     0.08     &    -128.1  $\pm$      1.9     &     0.464  $\pm$    0.021     &     0.883  $\pm$    0.086     &    -137.9  $\pm$      0.6  & -\\
 &   C4     &     0.023  $\pm$    0.004     &      0.95  $\pm$     0.08     &    -128.1  $\pm$      4.6     &     0.337  $\pm$    0.024     &     0.999  $\pm$    0.004     &     -16.4  $\pm$     30.8  & -\\
 &   C1     &     0.021  $\pm$    0.002     &      3.22  $\pm$     0.08     &    -131.6  $\pm$      1.4     &     0.765  $\pm$    0.028     &     0.999  $\pm$    0.008     &    -132.3  $\pm$      6.5  & -\\
 2016.850  & Core &     0.791  $\pm$    0.138     &      0.01  $\pm$     0.08     &      47.2  $\pm$    412.6     &     0.108  $\pm$    0.009     &     0.896  $\pm$    0.045     &    -142.4  $\pm$      0.8      &     7.99  $\pm$     2.01  \\
 &   C3     &     0.114  $\pm$    0.021     &      0.48  $\pm$     0.08     &    -136.1  $\pm$      9.5     &     0.106  $\pm$    0.040     &     0.993  $\pm$    0.028     &     -22.5  $\pm$     31.4  & -\\
 &   C2     &     0.026  $\pm$    0.004     &      2.31  $\pm$     0.08     &    -128.0  $\pm$      2.0     &     0.379  $\pm$    0.031     &     0.971  $\pm$    0.090     &    -140.7  $\pm$      0.7  & -\\
 &   C4     &     0.029  $\pm$    0.004     &      0.90  $\pm$     0.08     &    -126.4  $\pm$      5.4     &     0.469  $\pm$    0.057     &     0.959  $\pm$    0.112     &    -138.6  $\pm$      2.6  & -\\
 &   C1     &     0.022  $\pm$    0.002     &      3.18  $\pm$     0.08     &    -131.4  $\pm$      1.5     &     0.787  $\pm$    0.040     &     0.999  $\pm$    0.008     &    -134.2  $\pm$      3.1  & -\\
 2016.943  & Core &     0.801  $\pm$    0.135     &      0.01  $\pm$     0.10     &      41.6  $\pm$    496.9     &     0.106  $\pm$    0.005     &     0.997  $\pm$    0.017     &    -122.67  $\pm$      0.02      &     7.59  $\pm$     1.45  \\
 &   C3     &     0.081  $\pm$    0.013     &      0.67  $\pm$     0.10     &    -132.8  $\pm$      8.2     &     0.290  $\pm$    0.012     &     0.999  $\pm$    0.008     &    -149.1  $\pm$     37.0  & -\\
 &   C2     &     0.028  $\pm$    0.004     &      2.28  $\pm$     0.10     &    -127.8  $\pm$      2.5     &     0.471  $\pm$    0.042     &     0.994  $\pm$    0.034     &    -120.5  $\pm$     13.6  & -\\
 &   C1     &     0.016  $\pm$    0.003     &      3.30  $\pm$     0.11     &    -132.5  $\pm$      1.9     &     0.772  $\pm$    0.085     &     0.870  $\pm$    0.029     &    -144.0  $\pm$     14.5  & -\\
 2017.077  & Core &     0.791  $\pm$    0.137     &      0.01  $\pm$     0.09     &      46.9  $\pm$    415.3     &     0.118  $\pm$    0.002     &     0.999  $\pm$    0.004     &    -141.5  $\pm$      0.0      &     6.06  $\pm$     1.08  \\
 &   C3     &     0.096  $\pm$    0.017     &      0.52  $\pm$     0.09     &    -135.3  $\pm$      9.5     &     0.125  $\pm$    0.019     &     0.998  $\pm$    0.004     &     -22.1  $\pm$     31.5  & -\\
 &   C2     &     0.024  $\pm$    0.004     &      2.30  $\pm$     0.09     &    -128.0  $\pm$      2.3     &     0.398  $\pm$    0.073     &     0.999  $\pm$    0.003     &    -140.4  $\pm$      1.0  & -\\
 &   C4     &     0.018  $\pm$    0.003     &      1.05  $\pm$     0.09     &    -125.9  $\pm$      4.9     &     0.207  $\pm$    0.165     &     0.999  $\pm$    0.003     &     -20.5  $\pm$     35.1  & -\\
 &   C1     &     0.019  $\pm$    0.003     &      3.16  $\pm$     0.13     &    -131.4  $\pm$      2.4     &     0.787  $\pm$    0.132     &     0.999  $\pm$    0.002     &    -139.0  $\pm$     26.7  & -\\
 2018.307  & Core &     0.767  $\pm$    0.130     &      0.01  $\pm$     0.08     &      40.9  $\pm$    410.9     &     0.153  $\pm$    0.009     &     0.872  $\pm$    0.046     &     -49.0  $\pm$      4.2      &     3.99  $\pm$     0.86  \\
 &   C5     &     0.127  $\pm$    0.018     &      0.40  $\pm$     0.10     &    -129.2  $\pm$     14.1     &     0.487  $\pm$    0.066     &     0.998  $\pm$    0.008     &    -132.6  $\pm$      2.2  & -\\
 &   C4     &     0.020  $\pm$    0.003     &      2.30  $\pm$     0.08     &    -128.5  $\pm$      2.0     &     0.450  $\pm$    0.141     &     0.945  $\pm$    0.129     &    -139.2  $\pm$      8.5  & -\\
 &   C3     &     0.024  $\pm$    0.011     &      0.92  $\pm$     0.24     &    -121.3  $\pm$     13.3     &     0.683  $\pm$    0.374     &     0.945  $\pm$    0.190     &    -138.1  $\pm$     47.0  & -\\
 &   C2     &     0.014  $\pm$    0.003     &      3.27  $\pm$     0.09     &    -131.4  $\pm$      1.6     &     0.885  $\pm$    0.066     &     0.870  $\pm$    0.032     &    -162.1  $\pm$     15.4  & -\\
 2018.633  & Core &     0.555  $\pm$    0.115     &      0.04  $\pm$     0.07     &      54.6  $\pm$    112.8     &     0.035  $\pm$    0.092     &     1.000  $\pm$    0.002     &     -22.8  $\pm$     33.7      &      >    39.83  \\
 &   C6     &     0.243  $\pm$    0.075     &      0.12  $\pm$     0.08     &    -105.2  $\pm$     40.1     &     0.273  $\pm$    0.073     &     0.997  $\pm$    0.004     &    -136.6  $\pm$      1.8  & -\\
 &   C5     &     0.116  $\pm$    0.032     &      0.53  $\pm$     0.12     &    -130.2  $\pm$     12.0     &     0.702  $\pm$    0.035     &     0.997  $\pm$    0.006     &    -124.4  $\pm$      3.3  & -\\
 &   C4     &     0.021  $\pm$    0.003     &      2.30  $\pm$     0.07     &    -128.6  $\pm$      1.8     &     0.560  $\pm$    0.033     &     1.000  $\pm$    0.001     &     -13.9  $\pm$     28.1  & -\\
 &   C2     &     0.011  $\pm$    0.002     &      3.34  $\pm$     0.08     &    -131.8  $\pm$      1.3     &     0.765  $\pm$    0.059     &     1.000  $\pm$    0.001     &     -20.7  $\pm$     32.5  & -\\
 2018.863  & Core &     0.551  $\pm$    0.090     &      0.07  $\pm$     0.08     &      12.9  $\pm$     57.9     &     0.068  $\pm$    0.014     &     0.999  $\pm$    0.002     &     -17.3  $\pm$     28.0      &    12.54  $\pm$     5.58  \\
 &   C6     &     0.065  $\pm$    0.011     &      0.28  $\pm$     0.08     &    -114.8  $\pm$     16.4     &     0.007  $\pm$    0.021     &     0.999  $\pm$    0.004     &     -20.9  $\pm$     28.9  & -\\
 &   C5     &     0.090  $\pm$    0.011     &      0.64  $\pm$     0.08     &    -129.8  $\pm$      7.1     &     0.716  $\pm$    0.021     &     1.000  $\pm$    0.002     &    -116.7  $\pm$      1.7  & -\\
 &   C4     &     0.012  $\pm$    0.002     &      2.42  $\pm$     0.08     &    -129.0  $\pm$      1.8     &     0.261  $\pm$    0.035     &     0.999  $\pm$    0.003     &    -133.2  $\pm$      1.1  & -\\
 &   C2     &     0.012  $\pm$    0.001     &      3.15  $\pm$     0.08     &    -130.0  $\pm$      1.5     &     1.119  $\pm$    0.047     &     1.000  $\pm$    0.001     &    -120.1  $\pm$      4.3  & -\\
 2019.288  & Core &     0.518  $\pm$    0.091     &      0.02  $\pm$     0.08     &      32.9  $\pm$    277.4     &     0.108  $\pm$    0.068     &     0.879  $\pm$    0.054     &    -137.3  $\pm$      1.4      &     5.33  $\pm$     6.80  \\
 &   C6     &     0.057  $\pm$    0.022     &      0.47  $\pm$     0.13     &    -127.5  $\pm$     16.8     &     0.038  $\pm$    0.144     &     0.983  $\pm$    0.075     &     -20.8  $\pm$     30.4  & -\\
 &   C5     &     0.056  $\pm$    0.041     &      0.88  $\pm$     0.40     &    -129.0  $\pm$     23.8     &     0.709  $\pm$    0.509     &     0.938  $\pm$    0.243     &    -129.9  $\pm$     48.9  & -\\
 &   C4     &     0.021  $\pm$    0.003     &      2.64  $\pm$     0.12     &    -130.3  $\pm$      2.8     &     0.808  $\pm$    0.085     &     1.000  $\pm$    0.001     &    -125.4  $\pm$      2.8  & -\\
 2019.548  & Core &     0.748  $\pm$    0.159     &      0.00  $\pm$     0.11     &      58.4  $\pm$   1383.9     &     0.005  $\pm$    0.019     &     0.952  $\pm$    0.118     &     -33.2  $\pm$     38.7      &      >    30.71  \\
 &   C7     &     0.045  $\pm$    0.013     &      0.49  $\pm$     0.11     &    -130.9  $\pm$     13.0     &     0.139  $\pm$    0.516     &     0.985  $\pm$    0.061     &     -28.4  $\pm$     48.1  & -\\
 &   C5     &     0.022  $\pm$    0.010     &      1.00  $\pm$     0.24     &    -131.5  $\pm$     13.3     &     0.247  $\pm$    0.473     &     0.912  $\pm$    0.076     &     -60.9  $\pm$     17.3  & -\\
 &   C6     &     0.024  $\pm$    0.026     &      0.80  $\pm$     0.66     &    -114.1  $\pm$     28.7     &     1.100  $\pm$    0.862     &     0.883  $\pm$    0.094     &    -136.7  $\pm$     18.5  & -\\
 &   C4     &     0.008  $\pm$    0.004     &      2.31  $\pm$     0.24     &    -127.1  $\pm$      6.2     &     0.525  $\pm$    0.299     &     0.927  $\pm$    0.186     &     -37.6  $\pm$     73.3  & -\\
 &   C2     &     0.009  $\pm$    0.002     &      3.41  $\pm$     0.17     &    -131.7  $\pm$      2.9     &     0.784  $\pm$    0.311     &     0.993  $\pm$    0.031     &    -147.1  $\pm$      4.7  & -\\
 2020.399  & Core &     0.691  $\pm$    0.135     &      0.01  $\pm$     0.11     &      22.8  $\pm$    475.6     &     0.162  $\pm$    0.002     &     1.000  $\pm$    0.001     &      -9.0  $\pm$     34.8      &     2.78  $\pm$     0.55  \\
 &      U      &     0.100  $\pm$    0.016     &      0.52  $\pm$     0.11     &    -132.2  $\pm$     12.4     &     0.504  $\pm$    0.009     &     1.000  $\pm$    0.001     &     -26.1  $\pm$     44.8  & -\\
 &   C7     &     0.047  $\pm$    0.004     &      1.42  $\pm$     0.12     &    -128.2  $\pm$      4.6     &     1.502  $\pm$    0.026     &     1.000  $\pm$    0.001     &    -124.6  $\pm$      0.9  & -\\
 2020.585  & Core &     0.722  $\pm$    0.124     &      0.02  $\pm$     0.08     &      17.0  $\pm$    263.3     &     0.099  $\pm$    0.028     &     0.798  $\pm$    0.553     &      -1.5  $\pm$     13.2      &     9.82  $\pm$     8.97  \\
 &      U      &     0.152  $\pm$    0.034     &      0.38  $\pm$     0.08     &    -130.9  $\pm$     12.3     &     0.659  $\pm$    0.045     &     0.685  $\pm$    0.157     &    -123.1  $\pm$      1.4  & -\\
 &   C7     &     0.032  $\pm$    0.014     &      1.74  $\pm$     0.09     &    -125.5  $\pm$      3.0     &     1.338  $\pm$    0.052     &     0.436  $\pm$    0.004     &    -139.0  $\pm$      1.8  & -\\
 &   C4     &     0.008  $\pm$    0.007     &      3.33  $\pm$     0.28     &    -132.2  $\pm$      4.8     &     1.695  $\pm$    0.252     &     0.448  $\pm$    0.192     &    -130.2  $\pm$     15.5  & -\\
 2020.918  & Core &     0.568  $\pm$    0.094     &      0.02  $\pm$     0.08     &      37.4  $\pm$    300.5     &     0.155  $\pm$    0.005     &     0.877  $\pm$    0.060     &    -140.29  $\pm$      0.05      &     2.85  $\pm$     0.54  \\
 &      U      &     0.115  $\pm$    0.021     &      0.51  $\pm$     0.09     &    -127.2  $\pm$      9.9     &     0.758  $\pm$    0.047     &     0.867  $\pm$    0.017     &    -126.5  $\pm$      1.9  & -\\
 &   C7     &     0.023  $\pm$    0.003     &      2.06  $\pm$     0.09     &    -126.8  $\pm$      2.4     &     0.765  $\pm$    0.031     &     1.000  $\pm$    0.002     &    -128.1  $\pm$      1.0  & -\\
 
			 \bottomrule
		\end{longtable}
	
		\begin{longtable}{ccccccccc}
			\caption{CE model-fitting jet parameters for the 8-GHz map of PKS\,0502+049.} 	\label{table:Parameters_PKS_8GHz_table} \\
			\toprule
			Epoch & ID$^\mathrm{a}$ & $F$ & $r^\mathrm{b}$ & $\eta^\mathrm{c}$ & $a_\mathrm{FWHM}$ & Axial Ratio & $\mathrm{SPA}^\mathrm{c,d}$ & $T_\mathrm{B, rest}$ \\
			& & [Jy] & [mas] & [deg] & [mas] & & [deg] & $10^{11}\,\mathrm{K}$ \\
			\midrule
			\endfirsthead
			\multicolumn{9}{c}%
			{\tablename\ \thetable\ -- \textit{(Continued)}} \\
			\toprule
			Epoch & ID$^\mathrm{a}$ & $F$ & $r^\mathrm{b}$ & $\eta^\mathrm{c}$ & $a_\mathrm{FWHM}$ & Axial Ratio & $\mathrm{SPA}^\mathrm{c,d}$ & $T_\mathrm{B, rest}$ \\
			& & [Jy] & [mas] & [deg] & [mas] & & [deg] & $10^{11}\,\mathrm{K}$ \\
			\midrule
			\endhead
			\midrule \multicolumn{9}{r}{\textit{Continued on next page}} \\
			\endfoot
			\endlastfoot
 2018.841  & Core &     0.324  $\pm$    0.057     &      0.05  $\pm$     0.13     &      56.6  $\pm$    138.4     &     0.007  $\pm$    0.038     &     0.959  $\pm$    0.098     &     -21.8  $\pm$     28.6      &      >    13.25  \\
 &   C3     &     0.224  $\pm$    0.038     &      0.13  $\pm$     0.13     &    -127.0  $\pm$     60.6     &     0.600  $\pm$    0.052     &     0.998  $\pm$    0.008     &    -132.5  $\pm$      0.4  & -\\
 &   C2     &     0.033  $\pm$    0.012     &      0.68  $\pm$     0.14     &    -128.3  $\pm$     11.5     &     0.334  $\pm$    0.092     &     0.895  $\pm$    0.095     &      -9.0  $\pm$     24.0  & -\\
 &   C4     &     0.053  $\pm$    0.006     &      2.17  $\pm$     0.13     &    -127.2  $\pm$      3.5     &     1.135  $\pm$    0.028     &     0.999  $\pm$    0.005     &    -125.4  $\pm$      0.7  & -\\
 &   C1     &     0.009  $\pm$    0.001     &      4.13  $\pm$     0.16     &    -135.8  $\pm$      2.2     &     1.606  $\pm$    0.127     &     0.998  $\pm$    0.010     &     -14.9  $\pm$     29.2  & -\\

 	    \bottomrule
		\end{longtable}
	    \vspace{-10pt}
        \parbox{15.5cm}{\small \flushleft $^\mathrm{a}$ Here, core denotes the apparent origin of a jet where its optical depth of synchrotron emission reaches unity, C plus a number denotes the identified jet components and U means unidentified jet components.\\
	    $^\mathrm{b}$ Measured from the reference centre of the interferometric observations.\\
	    $^\mathrm{c}$ Measured in the direction of north to east. \\
	    $^\mathrm{d} \mathrm{SPA} = \psi - 90 ^\circ$.}
	\end{center}

\section{Cropping procedure and CE optimisations}\label{appendix:cropimgtest}

\begin{figure*}
	\centering
	\includegraphics[width=1.0\textwidth]{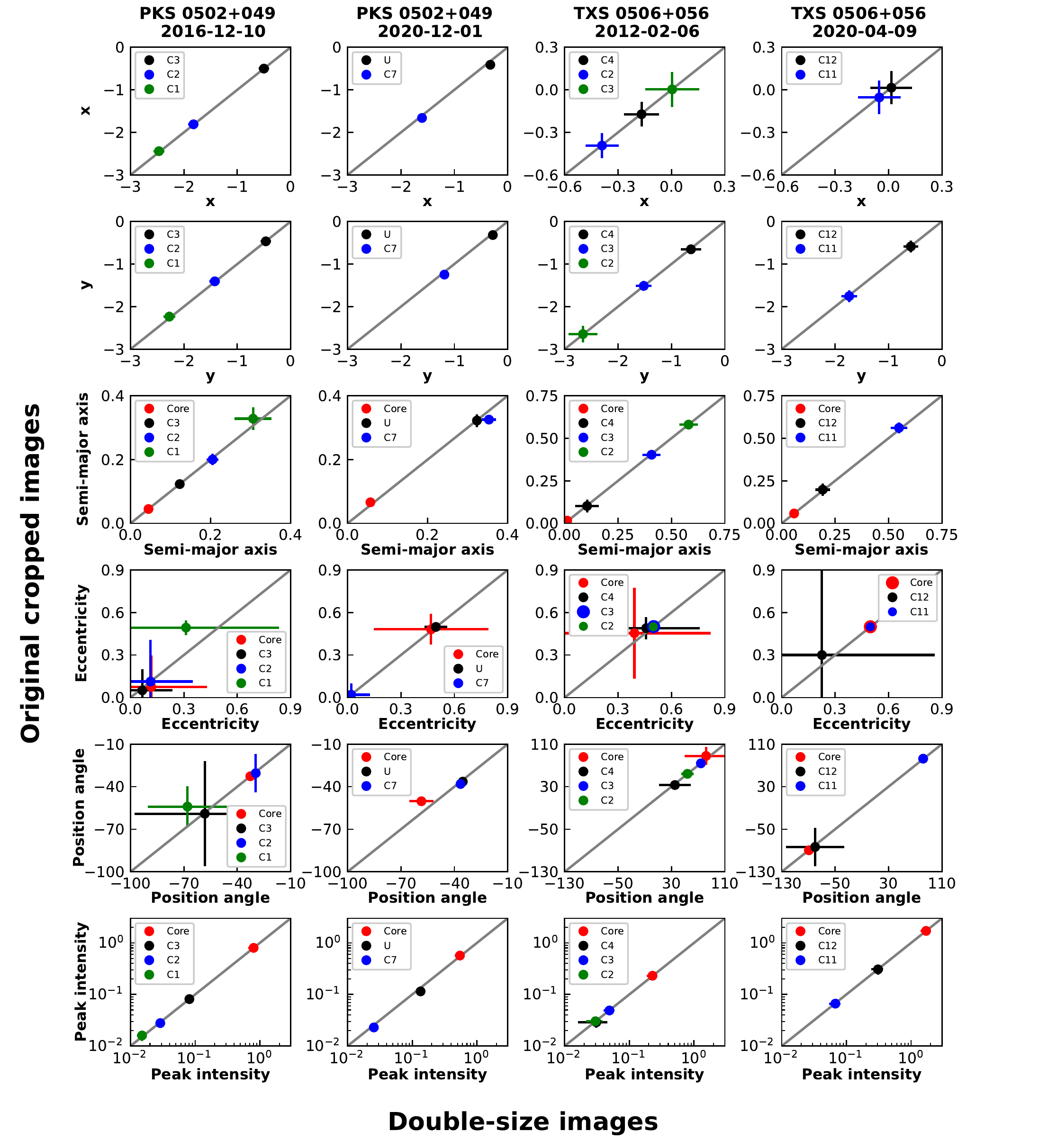}
    \caption{Comparison between the structural elliptical Gaussian parameters obtained from CE model-fittings for two images of PKS\,0502+049 (first two columns from left to right) and two images of TXS\,0506+056 (last two columns). Ordinate and abscissa axes correspond, respectively, to the values of the CE structural parameters obtained from the cropped images used in this work and from the new cropped images after doubling their original sizes. The Gaussian peak coordinates $x$ and $y$, as well as the semi-major axis are in units of milliarcseconds, while the structural position angle is given in degrees and the peak intensity (in logarithm scale) in units of Jy\,beam$^{-1}$. Error bars show the $1\sigma$ uncertainties of those quantities.}
    \label{fig:comparison_CE}
\end{figure*}

We show in \autoref{fig:comparison_CE} a comparison between structural parameters of the Gaussian components presented in Section \ref{sec:Results} and new CE optimisations applied to two double-size images of PKS\,0502+049 and TXS\,0506+056. These four representative epochs were randomly chosen among all the images analysed in this work. As already pointed out by \citet{Caproni_et_al_2014} who studied the parsec-scale jet of the quasar PKS\,1741-03, as well as in the case of CE validation tests performed by \citet{Caproni_et_al_2011}, no substantial differences (smaller than the involved uncertainties) were found among structural parameters of the jet components obtained in both situations. It reinforces the premise that cropping process does not interfere in the CE modellings of the images if all source's signal remains in the cropped image.

\bsp	
\label{lastpage}
\end{document}